%% file: main-aistats.tex
\documentclass[twoside]{article}

\input{preamble}

\usepackage{float}

\usepackage[accepted]{aistats2025}
\begin{document}



\twocolumn[

\aistatstitle{Transformers are Provably Optimal In-context Estimators for Wireless Communications}

\aistatsauthor{ Vishnu Teja Kunde, Vicram Rajagopalan, Chandra Shekhara Kaushik Valmeekam}
\aistatsauthor{Krishna Narayanan, Jean-Francois Chamberland,  Dileep Kalathil, Srinivas Shakkottai}
\aistatsaddress{ Texas A\&M University} 

% \aistatsauthor{ Vishnu Teja Kunde \And Vicram Rajagopalan}
% \aistatsaddress{ Texas A\&M University \And  Texas A\&M University} 
% \aistatsauthor{ Chandra Shekhara Kaushik Valmeekam \And  Krishna Narayanan }
% \aistatsaddress{ Texas A\&M University \And  Texas A\&M University} 
% \aistatsauthor{ Jean-Francois Chamberland  \And Dileep Kalathil \And Srinivas Shakkottai}
% \aistatsaddress{ Texas A\&M University \And  Texas A\&M University \And Texas A\&M University} 

]

\input{AISTATS-2025/contents-AISTATS/abstract-AISTATS}
\input{AISTATS-2025/contents-AISTATS/introduction}

\input{AISTATS-2025/contents-AISTATS/in-context-estimation}

\input{AISTATS-2025/contents-AISTATS/detection-loss}

\input{AISTATS-2025/contents-AISTATS/transformers-for-ICE}

\input{AISTATS-2025/contents-AISTATS/results-and-discussion}

\input{AISTATS-2025/contents-AISTATS/conclusion}

% \input{contents-AISTATS/softmax-attention-theory}

% \input{contents-AISTATS/softmax-for-BPSK}

\input{AISTATS-2025/contents-AISTATS/acknowledgements}

\bibliography{refs}
\bibliographystyle{apalike}

\input{AISTATS-2025/contents-AISTATS/checklist}

\appendix
\input{AISTATS-2025/contents-AISTATS/supplementary}

\end{document}

%% file: preamble.tex
\usepackage{tikz}
\usepackage[utf8]{inputenc}
\usepackage{amsmath, amssymb, bbm}
\usepackage[]{enumitem}

\usepackage{subcaption}

\usepackage{hyperref}

\newtheorem{theorem}{Theorem}

\newtheorem{lemma}[theorem]{Lemma}

\newtheorem{problem}{Problem}

\newtheorem{assumption}[problem]{Assumption}

\newtheorem{definition}[problem]{Definition}

\newcommand{\R}{\mathbb{R}}

\newcommand{\abs}[1]{\vert#1\vert}
\newcommand{\norm}[1]{\lVert#1\rVert}

\newcommand{\bC}{\mathbb{C}}

\newcommand{\bE}{\mathbb{E}}

\newcommand{\bH}{\mathbb{H}}
\newcommand{\bI}{\mathbb{I}}

\newcommand{\bP}{\mathbb{P}}

\newcommand{\bR}{\mathbb{R}}

\newcommand{\cC}{\mathcal{C}}

\newcommand{\cL}{\mathcal{L}}

\newcommand{\cN}{\mathcal{N}}

\newcommand{\cS}{\mathcal{S}}

\newcommand{\cU}{\mathcal{U}}

\newcommand{\cX}{\mathcal{X}}

\newcommand{\ve}{\mathbf{e}}

\newcommand{\vg}{\mathbf{g}}
\newcommand{\vh}{\mathbf{h}}

\newcommand{\vu}{\mathbf{u}}
\newcommand{\vv}{\mathbf{v}}

\newcommand{\vx}{\mathbf{x}}
\newcommand{\vy}{\mathbf{y}}
\newcommand{\vz}{\mathbf{z}}

\newcommand{\0}{\mathbf{0}}

\newcommand{\mC}{\mathbf{C}}

\newcommand{\mH}{\mathbf{H}}
\newcommand{\mI}{\mathbf{I}}

\newcommand{\mM}{\mathbf{M}}

\newcommand{\mR}{\mathbf{R}}

\newcommand{\mT}{\mathbf{T}}
\newcommand{\mU}{\mathbf{U}}

\newcommand{\mW}{\mathbf{W}}
\newcommand{\mX}{\mathbf{X}}

\newcommand{\bSigma}{\boldsymbol{\Sigma}}

%% file: AISTATS-2025/contents-AISTATS/abstract-AISTATS.tex
\begin{abstract}
 Pre-trained transformers exhibit the capability of adapting to new tasks through in-context learning (ICL), where they efficiently utilize a limited set of prompts without explicit model optimization.
 The canonical communication problem of estimating transmitted symbols from received observations can be modeled as an in-context learning problem: received observations are  a noisy function of transmitted symbols, and this function can be represented by an unknown parameter whose statistics depend on an  unknown latent context. This problem, which we term in-context estimation (ICE), has significantly greater complexity than the extensively studied linear regression problem.
 The optimal solution to the ICE problem is a non-linear function of the underlying context. In this paper, we prove that, for a subclass of such problems, a single-layer softmax attention transformer (SAT) computes the optimal solution of the above estimation problem in the limit of large prompt length. We also prove that the optimal configuration of such a transformer is indeed the minimizer of the corresponding training loss. Further, we empirically demonstrate the proficiency of multi-layer transformers in efficiently solving broader in-context estimation problems. Through extensive simulations, we show that solving ICE problems using transformers significantly outperforms standard approaches. Moreover, just with a few context examples, it achieves the same performance as an estimator with perfect knowledge of the latent context. The code is available \href{https://github.com/vishnutez/in-context-estimation}{here}.
\end{abstract}

%% file: AISTATS-2025/contents-AISTATS/introduction.tex
\section{INTRODUCTION}

 Pre-trained transformers \cite{vaswani2017attention} exhibit the capability of \textit{in-context learning} (ICL) - producing the correct output based only on the contextual examples provided in the prompt  \cite{brown2020langmodelsfewshot, xie2022explanation}. More precisely, when a pre-trained transformer is prompted with $k$ examples of input-output pairs $(\vx_{n},f(\vx_{n}, \theta))^{k-1}_{n=0}$ where $\theta$ is a context common to all the prompts, followed by an input query $\vx_{k}$, it can produce an output that is close to the ground truth $f(\vx_{k}, \theta)$ \cite{garg2023transformers}. Importantly, the function $f$ and context $\theta$ are sampled during the inference from a large function class $\mathcal{F}$ and context class $\Theta$, respectively, and these specific function and context may not have been in the data used for pre-training the transformer. Effectively, through ICL, a pre-trained transformer is able to learn a new function $f(\cdot, \theta)$ only using a small number of examples and without any explicit fine-tuning of its pre-trained parameters. This makes ICL a very useful property for many practical applications such as natural language processing \cite{min2022rethinking,min2022metaicl} and online decision making \cite{lee2024supervised, mukherjee2024pretraining}. Even in wireless communication, the efficiency of transformers for symbol estimation in various settings has been demonstrated empirically \cite{zecchin2024mimoicl, zecchin2024mimoeq}.

Inspired by the empirical investigation of ICL by \cite{garg2023transformers} from the perspective of learning particular function classes, many recent works have sought to develop a theoretical understanding of this phenomenon. In particular, many works have investigated the dynamics of ICL in transformers with a single linear self-attention layer on linear regression tasks \cite{zhang2024trained, wu2023many}. This style of theoretical analysis has also been extended to a single softmax self-attention layer \cite{huang2023incontextconvergencetransformers, collins2024context}, multi-headed attention layer  \cite{siyu2024training}, and to nonlinear function classes \cite{guo2023transformerslearnincontextsimple,yang2024incontextlearningrepresentationscontextual}. Another class of works has shown that each attention layer in a transformer can perform an implicit gradient descent step and explained the ICL phenomenon as an implicit gradient-based meta-learning \cite{von2023transformers, ahn2023transformers}. However, all these works focus on ICL for prediction tasks.  Due to page limitation, we defer a more detailed literature review to the Appendix.

% Although the interpretability of in-context learning for solving rich classes of problems can be very challenging, there have been attempts to understand the theoretical aspects of in-context learning for simpler problems, specifically linear regression. The work \cite{zhang2024trained} characterizes the convergence of the training dynamics of a single layer linear attention (LA) for solving a linear regression problem over random regression instances.
% It shows that this configuration of the trained transformer is optimal in the limit of large prompt length. 

Differently from the existing theoretical works mentioned before, in this paper, we introduce and investigate the \textbf{in-context estimation (ICE)} problem:  
\begin{quote}
    \textit{Prompted with $k$ examples of the form  $(g(\vx_n, \theta), \vx_n)^{k-1}_{n=0}$ where $g(\vx,\theta)$ is a stochastic function whose distribution is parameterized by $\theta$, followed by a query $g(\vx_{k}, \theta)$, can a pre-trained transformer correctly estimate the ground truth $\vx_{k}$ without any additional fine-tuning of its parameter?} 
\end{quote}

The above problem is an \textbf{inverse problem} \cite{Tarantola2005},\cite{bai2020dlforinvprobs}, which is significantly different from the forward prediction problems, such as linear regression, that have been addressed in the ICL literature. The key difficulty here lies in constructing an optimal estimator that depends on the noisy observations given in the prompt,  and the unknown prior distribution of the parameter $\theta$ (in Section \ref{sec:in-context-estimation}, we have given a more rigorous explanation on the challenges of solving inverse problems in-context and their difference from the forward prediction problem such as linear regression).

% the sequence $(g(\vx_1, \theta), \vx_1,\hdots,$ $g(\vx_k, \theta), \vx_k, g(\vx_{k+1}, \theta))$ where $g(\vx,\theta)$ is a stochastic function whose distribution is parameterized by $\theta$, and our goal is to estimate $\vx_{k+1}$. This inverse problem is inherently more challenging than predicting $f(\vx_{k+1}, \theta)$ for $\vx_{k+1}$ (as in linear regression) due to the complicated dependence of the optimal estimator on noisy observations.

In-context estimation can be a very useful property for many practical engineering applications such as deconvolution, denoising, and signal reconstruction. However, to keep the paper focused, we consider a subclass of in-context estimation problems that are of particular interest to the theory of (wireless) communication. Here, we consider  $\vx$ as a transmitted signal from a (wireless) communication transmitter, and a communication channel with the unknown state (context) $\theta$ maps $\vx$ into a received signal represented as $\vy = g(\vx,\theta)$. The receiver is then required to recover $\vx$ using the examples formed by the received symbols corresponding to the transmitted pilot symbols\footnote{Pilot symbols are pre-determined transmitted symbols known at the receiver.} (more details about the function $g(\vx,\theta)$ and the in-context examples are given  in Section~\ref{sec:in-context-estimation}). In standard communication systems, the received symbols and pilot symbols are first used to estimate the unknown channel parameters, and then this estimated channel parameter is used to estimate the unknown transmitted symbols. Both of these estimation processes typically involve solving an optimization problem in real-time, either using closed-form expressions under simplified model assumptions or by running an iterative procedure. This procedure can be particularly challenging when the channel parameter $\theta$ has a complex and sometimes unknown prior distribution. The goal of in-context estimation is to avoid this standard double estimation procedure by directly estimating the transmitted signal through a single inference step (forward pass) of a pre-trained transformer.

Our work is motivated by the observation that a pre-trained transformer's ability to perform in-context sequence completion suggests strongly that it should be able to approximate the desired posterior given the in-context examples for our inverse problem.  Essentially, if a transformer is pre-trained on a variety of contexts, it should be able to implicitly determine the latent context from pilot symbols and then perform end-to-end in-context estimation of transmitted symbols.  Once trained, such a transformer is simple to deploy because there are no runtime modifications, which could make it a potential building block of future wireless receivers. We indeed show this result both theoretically and empirically.

% In this paper, we study the in-context estimation (ICE) ability of transformers. 
% \item We make a fundamental connection between the in-context learning property of transformers with our desire for in-context estimation applicable to communication systems.
    %\item For special cases of such problem (the symbols ($\{\vx_t\}_1^{k+1}$) are of constant magnitude), we theoretically show that there exists a configuration of single layer {\bf softmax attention transformer (SAT)} that it is asymptotically optimal.
%We prove the existence of a single layer SAT which is an optimal estimator in the mean squared error sense given an asymptotically long prompt, for a subset of ICE problems.

Our main contributions are the following:
\begin{itemize}
    \item We formulate the inverse problem of estimating signals from noisy observations using in-context examples, which we call the in-context estimation problem. This is significantly different from the  (forward)  prediction problems like linear regression which is a well-studied in-context learning problem. So, the existing results cannot be directly used to analyze this problem. 
    \item For a special class of such problems (detailed in Section \ref{sec:main-results}) where the channel input-output relationship has a linear form given by $\vy = g(\vx,\theta) = \mH(\theta) \vx + \vz$, where $\mH(\theta)$ is an unknown channel parameter that depends on the unknown context $\theta$ with an unknown prior and $\vz$ is Gaussian noise, we show the following. \\
    $(i)$ There exists a single-layer softmax attention transformer that can perfectly estimate the true posterior of the transmitted signal in an asymptotic sense,  only using the examples given in the prompt and without performing any inference-time modification or finetuning of its pre-trained parameter. \\
    $(ii)$ The limit (with respect to the number of prompts in the training data and prompt length) of the cross-entropy loss function we use to train the single-layer softmax attention transformer is convex in the transformer parameters, after an appropriate parameterization of the original parameters. This essentially guarantees that gradient descent-based training can converge to the global minimum of this asymptotic loss function. In other words, we show that a trained single layer softmax attention transformer can perfectly estimate the posterior of the unknown transmitted signal in an asymptotic sense. 
    \item We empirically demonstrate that multi-layer transformers are capable of efficiently solving  ICE problems with finite in-context examples for multiple canonical problems in communication systems. We consider channel models that are more general than the i.i.d. linear models we considered for theoretical analysis. In all cases, we show that the in-context estimation performance of multi-layer transformers is similar or superior to the performance of the standard baseline algorithms, and sometimes even comparable to the performance of an oracle algorithm with full knowledge of the hidden channel parameter. 
\end{itemize}

%% file: AISTATS-2025/contents-AISTATS/in-context-estimation.tex
\section{PROBLEM FORMULATION: IN-CONTEXT ESTIMATION}

\textbf{Notation}. We denote random variables by capital letters (e.g. $X$), a vector as a boldface small letter (e.g. $\vx$), and a matrix with a boldface capital letter (e.g. $\mH$). We denote a set of first $S$ non-negative integers as  $[S] = \{0, \dots, S-1\}$. We denote $\ve_i$ to be the $i$th column of the identity matrix $\mI_d$ for $i\in[d]$.

\label{sec:in-context-estimation}
In this section, we define the problem of in-context estimation and discuss the challenges in solving it. We also discuss why this inverse problem is significantly different from the prediction (regression) problem in many settings. 
\begin{definition}
\label{def:ice}
{\bf (In-Context Estimation)} Let $\vx_{s_n} \in \cX \subset \bR^{d_\vx}$ be the transmitted signal which is a known deterministic embedding corresponding to the input symbol $s_n \in \mathcal{S}$ at index $n$, with $s_n  \stackrel{\rm iid}{\sim} \rho$, and $|\mathcal{S}| = |\mathcal{X}|$ is finite. The received signal $\vy_{n}$ at index $n$ is given by
\begin{align}
    \label{eq:ICE-model-1}
    \vy_{n} = \vg(\vx_{s_n}, \mH_n(\theta)) + \vz_n,
\end{align}
where $\mH_n(\theta) \sim f_{H \mid \theta}$   is a random process  determined by a latent parameter $\theta \sim f_\Theta$, and $ \vz_n \stackrel{\rm iid}{\sim}  f_Z$ is an additive noise. Let $(\vy_n, \vx_{s_n})_{n\in[N-1]}$ be the examples in the prompt and let $\vy_N$ be the query. The problem of in-context estimation (ICE) involves finding an estimate of the posterior for $s_{N}$ from the query $\vy_{N}$ and examples $\vy_{0:N-1}, \vx_{s_{0:N-1}}$ such that the probability of misclassification, $\bP[\hat{s}_{N} \neq s_{N}]$, is minimized.
\end{definition}

Note that since $\vx_{s_n}$ is a known deterministic embedding of the symbol $s_n$, having the examples $\vx_{s_{0:N-1}}$ is equivalent to having $s_{0:N-1}$.

\textbf{Optimal estimation:} The optimal estimate that minimizes the probability of misclassification  is based on the posterior, and is given by \cite{ProbTheoryOfPatternRecogBook}
\begin{align}
    \hat{s}_{N} = \arg\max_{i\in[S]} \bP[s_{N} = i \mid \vy_N, \vy_{0:N-1}, s_{0:N-1}].
    \label{eq:posterior-ideal-1}
\end{align}
Note that  $\mH_n(\theta)$ is unknown, and the above posterior computation requires the prior distribution over $\mH_n(\theta)$, which is typically unknown. In the simplified setting of a known prior, we can show that the posterior can be computed as 
\begin{align*}
    &\bP[s_{N} = i \mid \vy_N, \vy_{0:N-1}, s_{0:N-1}] \\
    &= c \rho_i \int_{\mH_{0:N}} f_{H_{0:N}}(\mH_{0:N}) d\mH_{0:N} \\
    & \cdot  \prod_{n\in[N]} f_{Y_n \mid H_n, S_n}(\vy_n \mid \mH_n, s_n) f_{Y_N \mid H_N, S_N}(\vy_N \mid \mH_N, i).
\end{align*}
Thus, even in the setting of a known prior, the above computation involves evaluating a multi-dimensional integral. In most practical estimation problems with a non-trivial prior, this computation can be very difficult or even intractable.  The ability of transformers to perform efficient in-context estimation circumvents the difficulty in the estimation of priors and computation of posteriors.

Motivated by canonical estimation problems in wireless communication, we restrict our attention to the problem of estimating complex symbols under an unknown transformation with additive  Gaussian noise, and we show that it naturally fits into the above framework.
In this class of problems, the relationship between the input symbols and the observed symbols (in the complex domain) is captured by 
\begin{equation}
    \label{eq:complex-est-prob}
    \tilde{\vy}_n = \tilde{\vh}_n(\theta) \tilde{x}_{s_n} + \tilde{\vz}_n,
\end{equation}
where $\tilde{x}_{s_n} \in \bC,  \tilde{\vh}_n(\theta), \tilde{\vz}_n, \tilde{\vy}_n \in \bC^d$. We can rewrite this complex-vector input-output representation as a real-matrix equation as
\begin{align}
\label{eq:real-matrix-eqns}
     \vy_n = \mH_n(\theta) \vx_{s_n} + \vz_n,
\end{align}
where
\begin{align}
   &\vy_n =
    \begin{bmatrix}
    {\rm Re}(\tilde{\vy}_n) \\
     {\rm Im}(\tilde{\vy}_n)
    \end{bmatrix}, \mH_n(\theta) = \begin{bmatrix}
        \operatorname{Re}(\tilde{\vh}_n(\theta)) & -\operatorname{Im}(\tilde{\vh}_n(\theta)) \\
        \operatorname{Im}(\tilde{\vh}_n(\theta)) & \operatorname{Re}(\tilde{\vh}_n(\theta))
    \end{bmatrix} \nonumber
     \\
    &  \vx_{s_n} = \begin{bmatrix}
        \operatorname{Re}(\tilde{x}_{s_n}) \\
        \operatorname{Im}(\tilde{x}_{s_n})
    \end{bmatrix},
    \vz_n =
     \begin{bmatrix}
         \operatorname{Re}(\tilde{\vz}_n) \\
         \operatorname{Im}(\tilde{\vz}_n)
     \end{bmatrix},  
     \label{eq:real-matrix-eqns-2}
\end{align}
$\vx_{s_n} \in \cX \subset \bR^2, \mH_n(\theta)\in \bR^{2d \times 2}, \vz_n, \vy_n \in \bR^{2d}$. Note that  \eqref{eq:real-matrix-eqns}  takes the form of our original formulation given in \eqref{eq:ICE-model-1}.

\textbf{Estimation vs prediction:} In order to clearly understand the challenges of the in-context estimation problem and to illustrate why it can be more challenging than the in-context prediction problem, we consider a simplified setting where the input-output relationship is linear. More precisely, in \eqref{eq:ICE-model-1}, we consider $\vg(\vx_{s_n}, \mH_n(\theta)) = \mH_n(\theta) \vx_{s_n}$, resulting in the input-output relationship $\vy_{n}  = \mH_n(\theta) \vx_{s_n} + \vz_n$.
Further, we consider the hypothetical case when $\mH_n(\theta)$ is known.

\textit{Prediction:} Here, the goal is to get the optimal estimate $\hat{\vy}_{n}$ of $\vy_{n}$ for a given $\vx_{s_n}$. The optimal estimate $\hat{\vy}_n$ obtained from observation $\vx_n$ and given $\mH$ is simply the linear transform $\hat{\vy}_n = \mH\vx_n$.

\textit{Estimation:} Here, the goal is to get the optimal estimate $\hat{s}_{n}$ corresponding to the transmitted signal $\vx_{s_n}$ for a given received signal ${\vy}_{n}$, which is an inverse problem.
The optimal estimate of $s_n$ depends on the prior distribution of $s_n$ and covariance of $\vz_n$.
In this paper, our interest is in the case where $s_n$ is uniformly distributed over a finite set.
The optimal estimate is no longer a simply linear transform of the given data ${\vy}_{n}$.

The main consequence of this difference is that recent results on using transformers for in-context prediction such as in \cite{garg2023transformers,zhang2024trained} do not extend directly to the estimation problem.

It should be emphasized that our interest in the paper is in the case when $\mH_n(\theta)$ is unknown and hence, the optimal estimates are more complicated both in the case of prediction and estimation.

%% file: AISTATS-2025/contents-AISTATS/detection-loss.tex
\section{ICE WITH A SINGLE LAYER SOFTMAX ATTENTION TRANSFORMER}
\label{sec:main-results}

% \vspace{-0.3cm}

In this section, we analyze the ICE problem given in Definition  \ref{def:ice} on the specific input-output relationship given in  \eqref{eq:complex-est-prob} and \eqref{eq:real-matrix-eqns}. We will focus only on a single-layer softmax attention transformer (SAT) and analyze its capability of performing ICE for our problem. We make the following assumption.

% In this section, we present our theoretical analysis for a single-layer softmax attention transformer (SAT) as an estimator for the problem corresponding to \eqref{eq:complex-est-prob}.
% We work with the following assumption.

% \vspace{-0.2cm}

\begin{assumption}
\label{assum:SIMO-model} In  \eqref{eq:complex-est-prob}: \\
$(i)$ The channel parameter  process is time invariant, i.e., $\tilde{\vh}_n(\theta) = \tilde{\vh}(\theta)$, where $\tilde{\vh}(\theta) \sim f_{H\mid \theta}$ is a \emph{fixed} hidden variable under the context $\theta \sim f_\Theta$. \\
$(ii)$ The input symbols $s_n \stackrel{\rm iid}{\sim} \rho$ where $\rho$ is some distribution  over $\mathcal{S}$.  The transmitted signal $\tilde{x}_{s_n} \in  \tilde{\cX}$ and $\tilde{\cX} \subset \bC$ is a finite subset of the unit circle in $\bC$, i.e., $\abs{z} = 1$ for any $z \in \tilde{\cX}$. \\
$(iii)$ The noise samples $\tilde{\vz}_n \stackrel{\rm iid}{\sim} \cC\cN(\0, \tilde{\bSigma}_{z})$ for some real positive definite matrix $\tilde{\bSigma}_z \in \bR^{d\times d}$.
% \begin{enumerate}[label=(\alph*)]
%     \item The hidden process is time invariant, i.e., $\tilde{\vh}_n(\theta) = \tilde{\vh}(\theta)$, where $\tilde{\vh}(\theta) \sim f_{H\mid \theta}$ is a \emph{fixed} hidden variable under the context $\theta \sim f_\Theta$.
%     \item The input indices $s_n \stackrel{\rm iid}{\sim} \rho$ where $\rho$ is some distribution on $S \triangleq \abs{\tilde{\cX}}$, and $\tilde{\cX} \subset \bC$ is a finite subset of the unit circle in $\bC$, i.e., $\abs{z} = 1$ for any $z \in \tilde{\cX}$. 
%     \item The noise samples $\tilde{\vz}_n \stackrel{\rm iid}{\sim} \cC\cN(\0, \tilde{\bSigma}_{z})$ for some real positive definite matrix $\tilde{\bSigma}_z \in \bR^{d\times d}$.
% \end{enumerate}
\end{assumption}

Under Assumption \ref{assum:SIMO-model}, we can show that the equivalent real-matrix equation \eqref{eq:real-matrix-eqns} satisfies the following properties: \\
$(i)$ The hidden channel parameter $\mH_n(\theta) = \mH(\theta)$ for all $n$, i.e., it is constant within a given prompt. \\
$(ii)$ For the transmitted symbol, $\norm{\vx_{s_n}}_2 = 1$ or all $n$. \\
$(iii)$ The noise $\vz_n \stackrel{\rm iid}{\sim} \cN(\0, \bSigma_z)$ such that  \\$\bSigma_z = \frac{1}{2} \begin{bmatrix}
    \tilde{\bSigma}_z & \0 \\
    \0 &  \tilde{\bSigma}_z
\end{bmatrix}$. 

From now on, we work with the above real quantities.
% Then, the real equations are given by
% \begin{equation}
% \label{eq:real-constant-h}
%     \vy_n = \mH(\theta) \vx_{s_n} + \vz_n, \quad n\ge 0
% \end{equation}
%  where $\vz_n \sim \cN(\0, \bSigma_z)$ such that $\bSigma_z \triangleq \frac{1}{2} \begin{bmatrix}
%     \tilde{\bSigma}_z & \0 \\
%     \0 &  \tilde{\bSigma}_z
% \end{bmatrix}$, and note that $\bSigma_z^{-1} = 2 \begin{bmatrix}
%     \tilde{\bSigma}_{z}^{-1} & \0 \\
%     \0 &  \tilde{\bSigma}_{z}^{-1}
% \end{bmatrix}$, and $\vx_{s_n} \in \cX$ can be identified as a subset of unit sphere $\bS^2$ in $\bR^2$. 
% Under the Assumption \ref{assum:SIMO-model}, $\mH_n = \mH$ for all $n$.

In the following lemma, we derive the true posterior of a transmitted symbol $s_q$ assuming that $\mH(\theta) = \mH$ is known. Note that when $\mH$ is known,  the posterior does not depend on $\theta$. Further, $(\vy_n, \vx_n )$ are conditionally i.i.d. for all $n$. Therefore, the posterior of $s_q$ can be computed by exploiting the   equation $\vy_q = \mH \vx_{s_q} + \vz_q$. 
\begin{lemma}
    \label{lem:true-posterior}
    Under Assumption \ref{assum:SIMO-model}, for $\vy_q = \mH \vx_{s_q} + \vz_q$, where $s_q \sim \rho$ and $\vz_q \sim \cN(\0, \bSigma_z)$, the posterior is given by
    \begin{align}
        &p_i(\vy_q, \mH; \bSigma^{-1}_z) = \bP[s_q = i \mid  \mH,  \vy_q] \nonumber \\
        &\hspace{1cm}= \frac{\rho_i\exp(\vy_q^T \bSigma^{-1}_z \mH \vx_{i})}{\sum_{j\in[S]} \rho_j \exp(\vy_q^T \bSigma^{-1}_z \mH \vx_{j})}, ~i\in[S].
        \label{eq:true-posterior}
    \end{align}
\end{lemma}

\emph{Proof}. Follows from elementary computation involving Bayes' theorem and Assumption \ref{assum:SIMO-model}. See Appendix for the full proof. $\hfill\square$

Note that, since $\mH(\theta)$ is not known in the problems we consider, \eqref{eq:true-posterior} can be thought of as an oracle posterior computation which provides a lower bound on the probability of error for any practical algorithm, including for our ICE approach.

Our goal is to train a single layer SAT for the in-context posterior estimation of the transmitted symbol using only the prompt that consists of $N$ in-context examples $\vy_{0:{N-1}}, s_{0:{N-1}}$ and the query $\vy_N$. More precisely, we want to investigate how well a single layer SAT can approximate $\bP[s_{N} = i \mid \vy_N, \vy_{0:{N-1}}, s_{0:{N-1}}]$. Towards, this, we first write down the input-output relationship of a single-layer SAT.

% Consider a single layer softmax attention transformer (SAT) to solve the in-context estimation of posterior of the query $s_q \triangleq s_{N}$ given observation $\vy_q \triangleq s_{N}$ and $N$ in-context examples $\vy_0^{N-1}, s_0^{N-1}$, where $\vy_q, \vx_{s_q}$ and $\{\vy_n, \vx_{s_n}\}_{n\in[N]}$ satisfy \eqref{eq:real-matrix-eqns}. 

Let $\vu_n$ denote the $n$th token in the input to the single layer SAT. We construct $\vu_n$ as  $\vu_n \triangleq [\vy_n^T~\ve_{s_n}^T]^T \in \bR^{2d+S}$ and $\vu_{N} \triangleq [\vy_q^T~\0_S^T]^T \in \bR^{2d+S}$.  Let $\tilde{\mW}_Q, \tilde{\mW}_K, \tilde{\mW}_V$ be the query, key, and value matrices of the single layer SAT. The attention scores between tokens $\vu_q$ and $\vu_n$ given by \cite{zhang2024trained} 
\begin{align}
\label{eq:attention-score-1}
    {\rm attn}_{q, n}(\tilde{\mW}_Q, \tilde{\mW}_K) = \frac{\exp(\vu_q^T \tilde{\mW}_Q^T \tilde{\mW}_K \vu_n)}{\sum_{m\in[N]}\exp(\vu_q^T \tilde{\mW}_Q^T \tilde{\mW}_K \vu_m)}.
\end{align}
And, the  $N$th output token ($q = N$)  is given by 
\begin{align}
    &\mT_{N}^{\rm SA}(\mU_{N}; \tilde{\mW}_Q, \tilde{\mW}_K, \tilde{\mW}_V) \nonumber \\
    &= \sum_{n\in[N]} \tilde{\mW}_V \vu_n  {\rm attn}_{q, n}(\tilde{\mW}_Q, \tilde{\mW}_K).
\end{align}

% \dk{cite-Bartlet paper, Sanjay-paper} \cite{collins2024context}
% \begin{align}
%     \mT_{N}^{\rm SA}(\mU_{N}) = \frac{\sum_{n\in[N]} \tilde{\mW}_V \vu_n \exp(\vu_{N}\tilde{\mW}_Q^T \tilde{\mW}_K \vu_n)}{\sum_{n\in[N]} \exp(\vu_{N}^T\tilde{\mW}_Q^T \tilde{\mW}_K \vu_n)}.
% \end{align}

% Let $\mT^{\rm SA}$ denote an SAT with the query, key, and value matrices $\tilde{\mW}_Q, \tilde{\mW}_K, \tilde{\mW}_V$ respectively, 

% acting on $(N+1)$ tokens $\vu_n \triangleq [\vy_n^T~\ve_{s_n}^T]^T \in \bR^{2d+S}$ for $n\in[N]$ and $\vu_{N} \triangleq [\vy_q^T~\0_S^T]^T \in \bR^{2d+S}$. Let $\mU_{N}$ be the matrix with columns $\vu_n$ for $n\in[N+1]$. Then, the $N$th output token (attention with previous tokens) is given by
% \begin{align}
%     \mT_{N}^{\rm SA}(\mU_{N}) = \frac{\sum_{n\in[N]} \tilde{\mW}_V \vu_n \exp(\vu_{N}\tilde{\mW}_Q^T \tilde{\mW}_K \vu_n)}{\sum_{n\in[N]} \exp(\vu_{N}^T\tilde{\mW}_Q^T \tilde{\mW}_K \vu_n)}.
% \end{align}

% The estimate of the above SAT for $\vx$ is provided by the last two entries in the $(n+1)$th output token, i.e., $[\mT_{n+1}^{\rm SA}]_{2d+1:2d+2}$. We denote this estimate with $n$ examples as $\hat{\vx}_{\rm SA}^n$.

It is straightforward to argue that the estimate of the SAT for the posterior of $s_{N}$, denoted by $p^N(\mU_N)$, can obtained from the last $S$ elements of the \(N\)th output token, i.e., \([\mT_{N}^{\rm SA}]_{2d:2d+S-1}\).
Thus, the first $N$ columns of $\tilde{\mW}_V$ do not affect the output $p^N(\mU_N)$, hence we set them to zeros without loss of generality. We note that many recent works on in-context regression have also considered similar parameterization of the query, key, and value matrices \cite{zhang2024trained, collins2024context}.  Motivated by these works, and  the form of the true posterior, we choose to re-parameterize the remaining entries of the weights as below:
\begin{align*}
    % \label{eq:query-key-values-2}
    \tilde{\mW}_Q &= \begin{bmatrix} \mW_Q & \0_{2d\times S} \\ \0_{S\times 2d} & \0_{S\times S}\end{bmatrix}, \quad \tilde{\mW}_K = \begin{bmatrix} \mW_K & \0_{2d\times S} \\ \0_{S\times 2d} & \0_{S\times S}\end{bmatrix}, \\
    \tilde{\mW}_V &= \begin{bmatrix}
    \0_{2d\times 2d} & \0_{2d\times S} \\ \0_{S\times 2d} & \mI_S
\end{bmatrix}.
\end{align*}

% \begin{align*}
%     \hat{\vx}^{\rm SA}_{n}(\mU_{n+1}; \mW) = \frac{\sum_{t=1}^n \vx_t \exp(\vy^T \mW \vy_t) }{\exp(\vy^T \mW \vy) + \sum_{t=1}^{n} \exp(\vy^T \mW \vy_t)}.
% \end{align*}
With these parameterizations, and by denoting $\mW = \mW_{\rm Q}^T \mW_{\rm K} \in \bR^{2d\times 2d}$,  the attention scores between tokens $\vu_q$ and $\vu_n$ given in \eqref{eq:attention-score-1} will reduce to
\begin{align*}
    {\rm attn}_{q, n}(\mW) &= \frac{\exp(\vy_q^T \mW \vy_n)}{\sum_{n\in[N]} \exp(\vy_q^T \mW \vy_n)}.
\end{align*}

The estimate of the SAT for the posterior of $s_{N}$, $p^N(\mU_N)$, can now be computed as 
\begin{align*}
    p^{N}(\mU_N; \mW) &= [\mT^{\rm SA}_{N}(\mU_N)]_{2d:2d+S-1}  \\
    &= \sum_{n\in [N]} \ve_{s_n} {\rm attn}_{q, n}(\mW).
\end{align*}
Thus, for $i\in[S]$, $p^N_i = \sum_{n\in [N]} \bI[s_n = i] {\rm attn}_{q, n} = \sum_{n\in \cS_i} {\rm attn}_{q, n}$, where the set $\cS_i \triangleq \{n\in[N]: s_n = i\}$, and $N_i \triangleq \abs{\cS_i}$, with $[N] = \sqcup_{i\in[S]} \cS_i$ and $N = \sum_{i\in[S]} N_i$. Hence, we obtain
\begin{align}
\label{eq:finite-posterior-sat}
    p^{N}_i(\mU_N;\mW) = \frac{ \sum_{n\in \cS_i}\exp(\vy_q^T \mW \vy_n) }{\sum_{i\in[S]}  \sum_{n\in \cS_i} \exp(
    \vy_q^T \mW \vy_n)}, 
\end{align}
which resembles a simple soft aggregation estimator. In the following lemma, we characterize the form of the above posterior estimate in the asymptotic setting. The proof is given in the Appendix and follows from an application of the strong law of large numbers.
\begin{lemma}
    \label{lem:convergence-SA}
For any $\theta \in \Theta$, suppose $\mH_n(\theta) = \mH$ is the common hidden parameter for $n \ge 0$, such that $\vy_n = \mH \vx_{s_n} + \vz_n$, and $\vy_q = \mH \vx_{s_q} + \vz_q$.
For a prompt $\mU_{N}$ with the $n$th column constructed as 
$\vu_n \triangleq [\vy_n^T, \ve_{s_n}^T]^T$ for $n\in[N]$ and $\vu_{N} \triangleq [\vy_q^T, \0_S^T]^T$, the estimate for the transformer with parameter $\mW$ satisfies a.s.,
\begin{align}
\label{eq:finite-posterior-sat-limit}
    \lim_{N \to \infty} p^{N}_i(\mU_N; \mW) = \frac{\rho_i \exp(\vy_q^T \mW \mH \vx_i)}{\sum_{j \in [S]}  \rho_j \exp(\vy_q^T \mW \mH \vx_j)}.
\end{align}  
\end{lemma}

% \emph{Proof}. Simplifying the estimate of the posterior made by the transformer as
% \begin{align*}
%      &p^{N}_i(\mW) = \frac{ \sum_{n\in \cS_i}\exp(\vy_q^T \mW \vy_n) }{\sum_{i\in[S]}  \sum_{n\in \cS_i} \exp(
%     \vy_q^T \mW \vy_n)}  \\
%     &=  \frac{ \left(\frac{N_i}{N}\right) e^{\vy_q^T \mW \mH \vx_{i}} \left(\frac{1}{N_i} \sum_{n\in \cS_i}e^{\vy_q^T \mW \vz_n} \right)}{\sum_{j\in[S]} \left(\frac{N_j}{N} \right) e^{\vy_q^T \mW \mH \vx_{j}}  \left ( \frac{1}{N_j}  \sum_{m\in \cS_j} e^{
%     \vy_q^T \mW \vz_m} \right)},
% \end{align*}
% the result follows by a careful application of the strong law of large numbers (see Appendix~\ref{app:detection-loss}). $\hfill\square$

We now give the first main result of our paper. 
\begin{theorem}
{\bf (Expressivity of SAT for optimal posterior estimation)} 
\label{thm:opt-SA-expressitivity}
%There exists a softmax attention transformer such that it is asymptotically an optimal estimator in the mean squared error sense, i.e., a.s.
There exists a softmax attention transformer such that,  its estimate converges to the true posterior as the prompt length increases to infinity, i.e., for $i\in[S]$
\begin{align*}
     \lim_{N \to \infty} p^N_i(\cdot; \bSigma_z^{-1}) = p_i( \cdot; \bSigma_z^{-1}) \quad \text{a.s.},
\end{align*}
where $\mU_N$ contains $s_n \sim \rho$ and $\vy_n$ with parameter $\mH$ and noise $\vz_{n} \sim \cN(\0, \bSigma_z)$.
\end{theorem}
\emph{Proof}. Using $\mW = \bSigma_z^{-1}$, this result follows directly from Lemma \ref{lem:true-posterior} and Lemma \ref{lem:convergence-SA}. $\hfill\square$

This establishes the fact that a single-layer softmax attention transformer is expressive enough to compute the posteriors and hence can achieve optimal performance in an asymptotic sense. 

We then turn to the training process: can the training of such a transformer with a large number of long prompts achieve optimal performance?

Denoting $Y_q, S_q, \{Y_n, S_n\}_{n\in[N]}$ as the random variables in the generation process, the empirical loss of the transformer for $T$ prompts of length $N$ is given by
\begin{align*}
    &\cL_{T, N}(\mW) \\
    &= -\frac{1}{T} \sum_{t\in[T]} \log p_{S_q(t)}^N(Y_q(t), Y_{0:N-1}(t), S_{0:N-1}(t); \mW).
\end{align*}
Asymptotic training loss of the transformer trained on a large number ($T\to \infty$) of long prompts ($N\to \infty$) is given by
\begin{align*}
    &\cL(\mW) \triangleq \lim_{T\to\infty} \lim_{N\to \infty} \cL_{T, N}(\mW) \\
    &= -\lim_{T\to\infty} \frac{1}{T} \sum_{t\in [T]} \log p_{S_q(t)}(Y_q(t), H(t); \mW),
\end{align*}
where the last equality uses the continuity of $\log$ and Theorem \ref{thm:opt-SA-expressitivity}.

Using the strong law of large numbers on the i.i.d. training prompts, we can simplify the above loss function as
\begin{align*}
    &\cL(\mW)= -\lim_{T\to\infty} \frac{1}{T} \sum_{t\in [T]} \log p_{S_q(t)}(Y_q(t), H(t); \mW) \\
    &=-\bE_{S_q, Y_q, H}\left[\log p_{S_q}( Y_q, H; \mW)\right] \\
    &= -\bE_{S_q, Y_q, H}\left[ \bE[\log p_{S_q}(Y_q, H; \mW) \mid Y_q, H]\right] \\
    &= -\bE_{S_q, Y_q, H} \left[ \bE_{S_q \sim p(Y_q, H; \bSigma^{-1})} \left[\log p_{S_q}( Y_q, H; \mW) \right] \right] \\
    &= -\bE_{S_q, Y_q, H} \left[ \sum_{i\in[S]} p_i( Y_q, H; \bSigma^{-1}) \log p_{i}(Y_q, H; \mW) \right].
\end{align*}
For characterizing $\cL$, we need the following lemmas, whose proofs are deferred to the Appendix.
\begin{lemma}
\label{lem:convexity-of-log-sum-exp}
The function $\ell: \bR^S \to \bR$, defined as $\ell(\vu) = \log \left(\sum_{i\in [S]} \exp(u_i)\right)$ is convex. If $\vu: V \to \bR^S$ is further a linear function, then the function $(\ell \circ \vu):V \to \bR$ is convex. 
\end{lemma}

\begin{theorem}
{\bf{(Convexity of the loss)}}
\label{thm:convexity-of-cross-entropy}
    The function $\cL: \bR^{d\times d} \to \bR_+$ is convex.
\end{theorem}
\emph{Proof}. It suffices to show that for a fixed $\vy_q, \mH$, the function $-\log p_i(\vy_q, \mH; \mW)$ is convex in $\mW$ for each $i\in [S]$. To this end, we observe that
\begin{align*}
   &-\log p_i(\vy_q, \mH; \mW) = -(\vy_q^T \mW \mH \vx_i + \log \rho_i) \\
   & \hspace{40 pt} + \log ( \sum_{j\in[S]} \exp(\vy_q^T \mW \mH\vx_j + \log \rho_j)).
\end{align*}
Now, noting that $u_j = \vy_q^T \mW \mH \vx_j$ is a linear function of $\mW$, the result follows from Lemma \ref{lem:convexity-of-log-sum-exp}.

\begin{theorem}
{\bf{(Global minimizer)}}
\label{thm:global-minimizer}
The global minimizer of $\cL$ is given by  $\mW^* = \bSigma_z^{-1}$.
\end{theorem}
\emph{Proof}. Noting that
\begin{align*}
    &-\bE_{S_q, Y_q, H}[\bE_{S_q \sim p(Y_q, H; \bSigma_z^{-1})}[\log p_{S_q}(Y_q, H; \bSigma_z^{-1})]] \\
    &\triangleq \bH[S_q \mid Y_q, H] 
\end{align*}
is the conditional entropy of $S_q$ given $Y_q$ and $H$, the Kullback-Leibler characterization of $\cL$ is given by
\begin{align*}
    &\cL(\mW) = \bH[S_q \mid Y_q, H] \\
    & + \sum_{i\in[S]} \bE_{S_q,Y_q,H}[D_{\rm KL}[p_{i}(Y_q,H; \bSigma_z^{-1}) \lVert p_i(Y_q, H; \mW)]].
\end{align*}
Thus, by non-negativity of KL-divergence, $\cL(\mW) \ge \bH[S_q \mid Y_q, H] = \cL(\bSigma_z^{-1})$ for any $\mW \in \bR^{d\times d}$, and so the result. $\hfill\square$

The above result says that when trained on a large number of long prompts, one can obtain the optimal configuration for SAT that computes the true posterior for our problem. \textit{In other words, we have shown that a trained single-layer softmax attention transformer can perfectly estimate the posterior of the unknown transmitted signal in an asymptotic sense. }

% \subsection{Relation between $\cL_N$ and $\cL$}

% {\color{red} VTK: Maybe we should just skip this section as the proof is neither complete nor clean}.

% In this section, we show that $\cL = \lim_{N\to \infty} \cL_N$. Note that $p_i^N(\mW) \in [0,1]$ is a random variable. Let $F^N_i(u) = \bP[p_i^N(\mW) \le u]$ be the distribution function. Note that $F^N_i(0^-) = 0$ and $F^N_i(1)=1$. By convergence of $p_i^N(\mW)$, we have that the distribution function for $p_i(\mW) > 0$ is $F_i(u) = 0$ for $u < p_i(\mW)$ and $F_i(u) = 1$ for $u \ge p_i(\mW)$. Thus, $F_i(0) = 0$. For any given $H$, the randomness in $p^N_i(\mW)$ is due to the partitions $\cS_i$ of $[N]$ and the noise $\{\vz_n\}_{n\in [N]}$. Thus, $\cL_N(\mW) = \sum_{i\in[S]} \bE[-\log p_i^N(\mW)]$, and so, we can write
% \begin{align*}
%     &\bE[-\log p_i^N(\mW)] = -\int_0^1 \log u ~d F^N_i(u) \\
%     &=[-(\log u) F_i^N(u)]\lvert_{u=0}^1 + \int_0^1 u^{-1} F_i^N(u) du \\
%     &= \int_0^1 u^{-1} F_i^N(u) du.
% \end{align*}
% Since $F_i^N(u) \downarrow 0$ for $u < p_i(\mW)$ and $F_i^N(u) \uparrow 1$ for $u \ge p_i(\mW)$ {\color{red} (hand-wavy)}, by the monotone convergence theorem and the continuity of $\log$, we get
% \begin{align*}
%    &\lim_{N\to \infty} \bE[-\log p_i^N(\mW) \mid Y_q, H] \\
%     &=  \bE\left[-\log \left(\lim_{N\to \infty} p_i^N(\mW) \right) \mid Y_q, H \right] \\
%     &= -\log p_i(\mW; Y_q, H). 
% \end{align*}

%% file: AISTATS-2025/contents-AISTATS/transformers-for-ICE.tex
\section{EXPERIMENTS WITH MULTI-LAYER TRANSFORMERS}
\label{in_context_estimation}

\subsection{Experimental setup}

We study the performance of multi-layer transformers empirically through the following experimental setup, which is inspired by common communication systems.
We choose the input space to be $\tilde{\cX} \triangleq \{\pm 1 \pm i: i\triangleq \sqrt{-1}\}$, which is a finite subset of the circle of radius $\sqrt{2}$ in $\bC$.
We assume that these symbols are equally likely, $\rho_i = \frac{1}{|\tilde{\cX}|}$.
The noise is distributed as $f_Z = \cC\cN(\0, 2\sigma^2 \mI_d)$. The quantity $1/\sigma^2$ is called the signal-to-noise ratio (SNR).

We consider two scenarios depending on the nature of the hidden parameter process $\tilde{\vh}_t(\theta)$.

% \subsubsection{Scenario 1: Time invariant process}

\textbf{Scenario 1: Time invariant process:} The setup is similar to that of our theoretical results. The hidden context $\theta$ can be either $0$ or $1$, i.e., $\Theta \triangleq \{0,1\}$. We consider $f_\Theta = \cU(\Theta)$, the uniform distribution on $\Theta$. The distributions $f_{H \mid \theta}$ are described next for $\theta \in \Theta$.
For $\theta = 0$, the $j$th component of $\tilde{\vh}(0)$ is given as $\tilde{h}^j(0) = \exp \left( \frac{-i\pi(j-1)\cos(\alpha)}{2} \right) \triangleq  \tilde{h}^j_\alpha(0), 
   j\in [d], \quad \alpha \sim \cU((0, \pi]),$ 
   % \begin{align*}
%    &\tilde{h}^j(0) = \exp \left( \frac{-i\pi(j-1)\cos(\alpha)}{2} \right) \triangleq  \tilde{h}^j_\alpha(0), \\
%    &j\in [d], \quad \alpha \sim \cU((0, \pi]),
% \end{align*}
where the subscript denotes the dependence on $\alpha$. The above distribution corresponds to a one-ray line-of-sight channel in wireless communication \cite{goldsmith_2005}, where $\alpha$ is the angle of arrival of the incoming signal at the receiver, and the antenna spacing is $(1/4)$th of the wavelength of the carrier.
When $\theta = 1$, we have $\tilde{\vh}(1) \sim \cC\cN(\0, \mI_d)$.
% \begin{align*}
%     \tilde{\vh}(1) \sim \cC\cN(\0, \mI_d).
% \end{align*}
This is referred to as the independent and identically distributed (i.i.d) Rayleigh fading channel model in the wireless communication literature \cite{goldsmith_2005}.

\textbf{Scenario 2: Time varying process:} Let latent space $\Theta$ be a compact subset of $\bR_+$, and let $f_{\Theta}$ be the uniform distribution over $\Theta$.
Further, assume $\{h^j_n(\theta)\}_{n\ge 0}$ are iid zero-mean wide-sense stationary (WSS) Gaussian processes (Section 15.5, \cite{Kay97}) across $j\in[d]$ with $\bE \left[ h^j_n(\theta) h^j_{n+k}(\theta) \right] \triangleq R_\theta(k) = J_0 \left( \frac{2\pi f_{\rm carrier} T_s k \theta}{c} \right),~k\ge 0$,
% \begin{align*}
%     \bE \left[ h^j_n(\theta) h^j_{n+k}(\theta) \right] &\triangleq R_\theta(k) \\
%     &= J_0 \left( \frac{2\pi f_{\rm carrier} T_s k \theta}{c} \right),~k\ge 0
% \end{align*}
where $J_0 (\cdot)$ is the Bessel function of the first kind of order zero. The above distribution is referred to as Clarke's model (Section 3.2.1, \cite{goldsmith_2005}) for time-varying channel processes in wireless communication. The constants $T_s, f_{\rm carrier}, c$ denote the symbol duration, carrier frequency, and the velocity of light respectively. The latent context $\theta$ denotes the relative velocity between the transmitter and the receiver.

% \subsection{Training}
% \label{sec:training}

We adopt the training setup described in \cite{panwar2024incontext}, which is in turn based on the approach used in \cite{garg2023transformers} with minor changes to account for complex-valued symbols. 
For all experiments, we train a GPT-2 \cite{radford2019language} decoder-only model from scratch.  The model consists of $12$ layers, with $8$ heads, and an embedding dimension of $256$. We have included our model architecture details, other hyper-parameter settings, and detailed training procedures in the Appendix.

In all the experiments, we set $d = 4$, and the constants $f_{\rm carrier} = 2.9\times 10^9~{\rm Hz}, T_s = 1~{\rm ms}, c = 3\times 10^8~{\rm m/s}$. For scenario 2, we consider $\Theta = \{5, 15, 30\}$. These are representative of the typical values seen in practice for wireless communications.

\vspace{-0.3cm}
\subsection{Baselines}
\label{sec:baselines}

We now describe some feasible context-aware posterior estimators that achieve low cross entropy, along with some context-unaware posterior estimators for both scenarios, motivated by wireless communications problems. More details about these baselines are given in the Appendix. 

%$\hat{\vx}^{\rm CA/CME, \theta}$
%$\hat{\vx}^{\rm CU/CME/LMMSE}$

% lower-bound that computes the true posterior on the symbol given the true context $\bP[S_q \mid \vy_0^n, s_0^{n-1}, \theta]$ (see \eqref{eq:scenario-1-aware} in Appendix) denoted as CA-Post.

\textbf{CA-Post:} For scenario 1, we provide a \textit{context-aware posterior (CA-Post)} $\bP[S_q \mid \vy_{0:n}, s_{0:n-1}, \theta]$ assuming that the context $\theta$ is known.  This is not a practical solution as $\theta$ is unknown. We use this as a lower bound to compare the performance of other algorithms. 

\textbf{CU-Post-H-LMMSE:}  For scenario 1, we also provide a context-unaware baseline that performs a Linear Minimum Mean Square Estimate (LMMSE) of the hidden parameter $\tilde{\vh}(\theta)$, and uses this estimate to perform a posterior on $S_n$  as $\bP[S_n \mid H_n={\rm LMMSE}[H_n \mid \vy_{0:n-1}, s_{0:n-1}]]$. This baseline is typically used in wireless communications.

% (see Appendix \ref{appen:typical-baselines}) and is denoted by CU-Post-H-LMMSE. Both these baselines require the knowledge of the statistics of the hidden parameter $\tilde{\vh}(\theta)$, but not the true latent parameter $\theta$.

%or $\hat{\vx}^{\rm CA/CME/MMSE}$
%or $\hat{\vx}^{\rm CU/CME/MMSE}$
%or $\hat{\vx}^{\rm CU/CME}$

% For scenario 2, we provide four ways of computing the posterior, a context-aware lower bound and three context-unaware baselines. The context-aware lower-bound computes the true posterior of the symbol given the true context $\theta$ as $\bP[S_n \mid \vy_0^{n}, s_0^{n-1}, \theta]$, denoted by CA-Post. This is infeasible in practice as the true context $\theta$ is unknown. 

% The first context-unaware baseline CU-Post-H-LMMSE computes the posterior $\bP[S_n \mid \vy_0^{n}, s_0^{n-1}, H_n]$ over $S_n$ by setting $H_n = {\rm LMMSE}[H_n \mid \vy_0^{n-1}, s_0^{n-1}]]$ which is similar to the one in scenario 1, albeit as the distributions are different in each scenario the computation of the quantities would vastly differ. 

\textbf{CU-Post-H-MMSE:} This context-unaware baseline computes an MMSE on ${\tilde{\vh}_t}$ (without  knowledge of $\theta$) and uses this estimate $H_n = {\rm MMSE}[H_n \mid \vy_{0:n-1}, s_{0:n-1}]$, to compute the posterior on the symbol $S_n$. This is computationally intensive.

\textbf{CU-Post:} This context-unaware baseline  computes the posterior on the symbol $\bP[S_n \mid \vy_{0:n}, s_{0:n-1}]$ without any knowledge of $\theta$, denoted by \emph{CU-Post}. This is the Bayesian estimate of the posterior and it is typically computationally very intensive.

%% file: AISTATS-2025/contents-AISTATS/results-and-discussion.tex
\subsection{Experiment Results and Discussion}

% mutual information is defined as
% \begin{align}
%     &{\rm MI}(p^N) = \bH[S_q] - \bH[p^N(S_q \mid Y_q, Y_{0:N-1}, S_{0:N-1})] \nonumber \\
%     &= \log S + \bE[\log p^N(S_q \mid Y_q, Y_{0:N-1}, S_{0:N-1})] \label{eq:MI},
% \end{align}
% where the expectation is over all the randomness in the prompt. We remark that this metric is a measure of quality of the posterior approximation, and should not be confused with the information-theoretic mutual information which is known to be non-negative. We choose this metric as opposed to just the average accuracy $\bE[\bP[\hat{S_q}\neq S_q]]$ of the maximum aposteriori estimator based on $p^N$, since computing a good approximation to the posterior can be used in downstream tasks.  

% Next, we present experimental results for scenarios 1 and 2. In short, across the scenarios we described, we show that GPT-2 models trained on sequences sampled from these scenarios are able to approach the performance of context-aware estimates. This is surprising, as the latter benchmark estimators are given the exact realization of $\theta$ for each case, while the models have to implicitly estimate this from the context in addition to predicting the symbols $\vx$.

We present experimental results for scenarios 1 and 2. The transformer, along with the other methods, is used to estimate the posterior over the transmitted signal. The quality of the posterior is evaluated using the cross-entropy with the true transmitted signal index, the lesser the cross-entropy, the better the posterior estimate. The key empirical result is that, across the scenarios we described, the GPT-2 models trained on sequences sampled from these scenarios are able to approach the performance of context-aware estimates. This is surprising, as the latter benchmark estimators are given the exact realization of $\theta$ for each case, while the models have to implicitly estimate this from the context in the processing of estimating the posterior over the symbols $s_q$.

\begin{figure*}[!ht]
     \centering
        \includegraphics[width=0.8\textwidth]{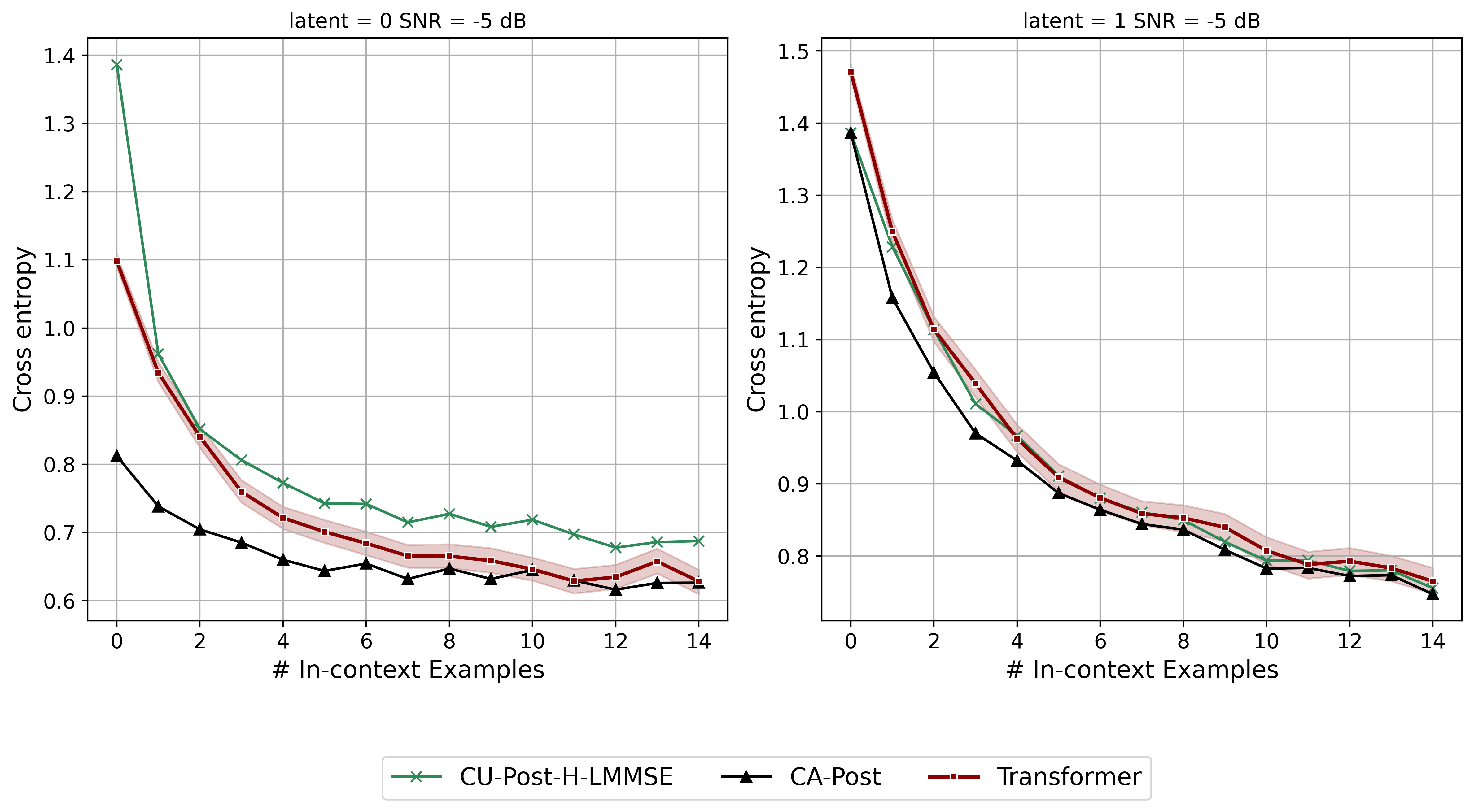}
    \caption{\small Cross entropy in scenario 1 as a function of number of in-context examples. The performance of the transformer (red) is close to that of the optimal context-aware estimator (black)  while significantly outperforming a typical baseline (green) when latent context $\theta=0$.}
    \label{fig:time_invariant_detection_snr_neg5}
\end{figure*}

\textbf{Scenario 1: Time invariant process:}   In Figure~\ref{fig:time_invariant_detection_snr_neg5}, for $\theta=0$, even the zero-shot performance of the transformer model is better than the context-unaware model. This indicates that the model is able to use the statistics of the $\vy$ vectors to distinguish between $\theta=0$ and $\theta=1$ without any explicit in-context example. As the context length increases, for both SNR values, the transformer model performs better than the context-unaware baseline and approaches the performance of the context-aware estimator.

% %%%%%%%%%%%%%%%%%%%%%%%%% main plot below %%%%%%%%%%%%%%
% \begin{figure*}
%     \centering
%     \includegraphics[width=\textwidth]{figures/scenario1-combined-low-dpi.png}
%     \caption{\small Estimation error in scenario 1 as a function of a number of in-context examples. The performance of the transformer (red) is close to that of the optimal context-aware estimator (black)  while significantly outperforming a typical baseline (green) when latent context $\theta=0$.}
%     \label{fig:ray_fading_results}
% \end{figure*}
% %%%%%%%%%%%%%%%%%%%%%%%%% main plot above %%%%%%%%%%%%%%  

% This shows that the model is able to determine the latent context and make a good estimate of the channel parameters.

When $\theta=1$, the three methods perform roughly the same across all context lengths. This is to be expected. The hidden parameter $\tilde{\vh}(1)$ in this case is Gaussian, and simply estimating the mean and variance of the channel will grant the best performance. The notable aspect of this inquiry is that the transformer model is able to correctly identify the context and has learned the correct way to estimate the channel for this context.

\begin{figure*}[!ht]
     \centering
    \includegraphics[width=\textwidth]{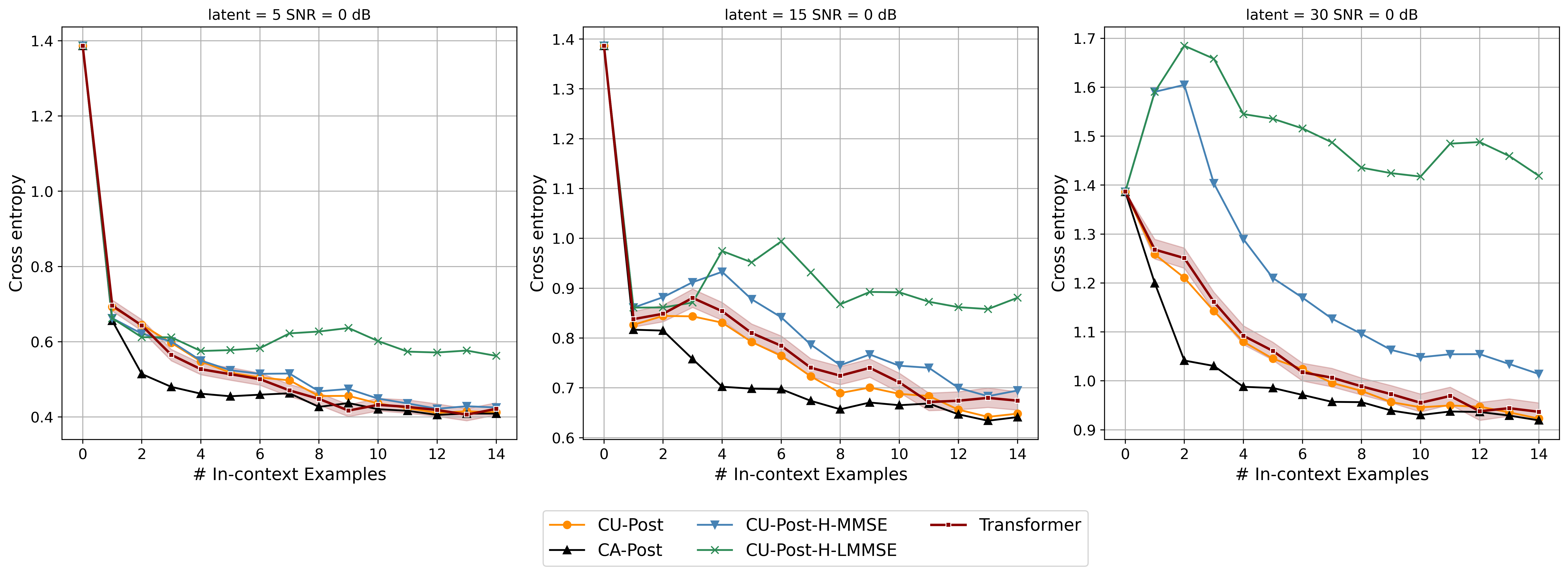}
    \caption{\small Cross entropy in scenario 2 as a function of number of in-context examples. The transformer (red) computes the optimal estimator (orange) with its performance approaching the context-aware estimator (black) as the number of examples increases, while significantly outperforming a typical baseline (green) for all the latent contexts.}
    \label{fig:time_varying_detection}
\end{figure*}

\textbf{Scenario 2: Time varying process:}  We present results for this scenario in Figure~\ref{fig:time_varying_detection}. These plots are produced in a manner similar to Figure~\ref{fig:time_invariant_detection_snr_neg5}, with the added challenge that the hidden parameters $\tilde{\vh}_n(\theta)$ vary with time $n$.

% %%%%%%%%%%%%%%%%%%%%%%%%% main plot below %%%%%%%%%%%%%%
% \begin{figure*}
%     \centering
%     \includegraphics[width=\textwidth]{figures/scenario2-combined-SNR0-low-dpi.png}
%     \caption{\small Estimation error in scenario 2 as a function of number of in-context examples. The transformer (red) computes the optimal estimator (orange) with its performance approaching the context-aware estimator (black) as the number of examples increases, while significantly outperforming typical baselines (green, blue) for latent contexts $\theta=15, 30$.}
%     \label{fig:time_varying}
% \end{figure*}
% %%%%%%%%%%%%%%%%%%%%%%%%% main plot above %%%%%%%%%%%%%%

% In Figure~\ref{fig:time_varying_detection}, we can see that the transformer model and other estimators have an accuracy of $25\%$ with zero examples for every $\theta$, which is as good as random guessing (since $S=4$).
% Because the distribution of the noise $\vz_1$ and the parameter $\tilde{\vh}_1(\theta)$ is symmetric about $\0$, any model requires at least one $(\vy, \vx)$ pair before it is able to make a meaningful prediction of the symbol. 

In Figure~\ref{fig:time_varying_detection}, across all $\theta$, the performance of the transformer-based estimator matches the performance of the context-unaware estimator CU-Post at every context length. Although we could compute CU-Post explicitly for this scenario, for general distributions of the hidden process, the computation is intractable. For this setting, our result provides strong evidence of the optimality of transformers in in-context estimation of the posterior.

This indicates that, with enough examples, the model is able to estimate the latent context $\theta$ and the hidden parameters $\tilde{\vh}_n(\theta)$, although the process is time-varying. There is a significant gap between the performance of the context-unaware baseline methods that make a point estimate of the hidden process $H_n$ CU-Post-H-LMMSE and CU-Post-H-MMSE, and the performance of the other methods, demonstrating that performing well requires estimating the posterior distribution very well. The transformer model is able to accurately infer the context and use this information to perform nearly as well as the context-aware estimator.

% As the context length increases both these estimates approach the performance of the context-aware estimator. 

A typical communication would include an error correction code and a decoder. The posterior probabilities are used in a decoder for decoding the code. Therefore, measuring the quality of posterior estimation (through cross-entropy) is more meaningful than just the raw accuracy of symbol prediction (we provide accuracy plots in Appendix \ref{appen:accuracy-plots}).

% The results shown here refer to the accuracy before decoding. If the rate of the code is chosen appropriately, the eventual accuracy can be made arbitrarily close to $100\%$.

%% file: AISTATS-2025/contents-AISTATS/conclusion.tex
\section{CONCLUSION}

In conclusion, our research demonstrates that pre-trained transformers can effectively adapt to new tasks through in-context learning (ICL), leveraging a limited set of prompts without explicit model optimization. We framed the canonical communication problem of estimating transmitted symbols from received observations as an in-context learning problem, which we termed in-context estimation (ICE). 

% This problem involves a noisy function of transmitted symbols represented by an unknown parameter whose statistics depend on an unknown latent context, presenting a complexity significantly greater than the well-studied linear regression problem due to its non-linear optimal solution.

Our findings show that, for a subclass of ICE problems, a single-layer softmax attention transformer (SAT) can compute the optimal solution in the limit of large prompt lengths. We provided proof that the optimal configuration of such a transformer is indeed the minimizer of the corresponding training loss and that the loss is convex. Additionally, our empirical results highlight the efficiency of multi-layer transformers in addressing broader ICE problems.
Simulations confirm that transformers not only outperform standard approaches but also match the performance of an estimator with perfect knowledge of the latent context using only a few context examples. These results underscore the potential of transformers in efficiently solving complex estimation problems in communication systems.

% Across the various tasks, parameters, and regimes studied, transformer models are able to infer the latent context from a few samples and then use this information to make an accurate estimation of $\vx_{k+1}$. We use genie-aided context-aware estimators to find lower bounds on the performance of any such model. These estimators are given exact, noiseless information about the context, which is not available to the transformer model. Even without such extra information, the transformer models we trained are able to approach and reach the performance of the context-aware methods. This finding demonstrates that the ability of transformers to learn in-context from prompts enables these models to perform well on wireless communication problems.

%% file: AISTATS-2025/contents-AISTATS/acknowledgements.tex
\subsection*{Acknowledgements}

We thank the department of Electrical and Computer Engineering (ECEN) at Texas A\&M University for providing access to the Olympus computing cluster. Portions of this research were conducted with the advanced computing resources provided by Texas A\&M High Performance Research Computing (HPRC). This work was supported in part by the National Science Foundation (NSF) grants NSF-CNS 2312978, NSF-CAREER-EPCN-2045783 and CNS-526050-00002. 
This work was also supported in part by the National Science Foundation grant NSF-CNS 2148354, federal agencies, and industry partners as specified in the Resilient \& Intelligent NextG Systems (RINGS) program.
Any opinions, findings, and conclusions or recommendations expressed in this material are those of the authors and do not necessarily reflect the views of the sponsoring agencies.

%% file: AISTATS-2025/contents-AISTATS/checklist.tex
\section*{Checklist}

 \begin{enumerate}

 \item For all models and algorithms presented, check if you include:
 \begin{enumerate}
   \item A clear description of the mathematical setting, assumptions, algorithm, and/or model. [Yes]
   \item An analysis of the properties and complexity (time, space, sample size) of any algorithm. [Yes]
   \item (Optional) Anonymized source code, with specification of all dependencies, including external libraries. [Yes]
 \end{enumerate}

 \item For any theoretical claim, check if you include:
 \begin{enumerate}
   \item Statements of the full set of assumptions of all theoretical results. [Yes]
   \item Complete proofs of all theoretical results. [Yes]
   \item Clear explanations of any assumptions. [Yes]   
 \end{enumerate}

 \item For all figures and tables that present empirical results, check if you include:
 \begin{enumerate}
   \item The code, data, and instructions needed to reproduce the main experimental results (either in the supplemental material or as a URL). [Yes]
   \item All the training details (e.g., data splits, hyperparameters, how they were chosen). [Yes]
         \item A clear definition of the specific measure or statistics and error bars (e.g., with respect to the random seed after running experiments multiple times). [Yes]
         \item A description of the computing infrastructure used. (e.g., type of GPUs, internal cluster, or cloud provider). [Yes]
 \end{enumerate}

 \item If you are using existing assets (e.g., code, data, models) or curating/releasing new assets, check if you include:
 \begin{enumerate}
   \item Citations of the creator If your work uses existing assets. [Yes]
   \item The license information of the assets, if applicable. [Yes]
   \item New assets either in the supplemental material or as a URL, if applicable. [Yes]
   \item Information about consent from data providers/curators. [Yes]
   \item Discussion of sensible content if applicable, e.g., personally identifiable information or offensive content. [Not Applicable]
 \end{enumerate}

 \item If you used crowdsourcing or conducted research with human subjects, check if you include:
 \begin{enumerate}
   \item The full text of instructions given to participants and screenshots. Not Applicable.
   \item Descriptions of potential participant risks, with links to Institutional Review Board (IRB) approvals if applicable. [Not Applicable]
   \item The estimated hourly wage paid to participants and the total amount spent on participant compensation. [Not Applicable]
 \end{enumerate}

 \end{enumerate}

%% file: AISTATS-2025/contents-AISTATS/supplementary.tex
\onecolumn

% {\hfill \Large \bf \underline{ ~Supplementary Materials} \hfill}

\input{AISTATS-2025/contents-AISTATS/additional-related-work}

\section{Proofs of the theoretical results}

\label{app:detection-loss}

\emph{Proof of Lemma \ref{lem:true-posterior}}. Using Bayes' theorem, we have
\begin{align*}
    &\bP[S_q = i \mid H = \mH, Y_q = \vy_q] \\
    &= c_0\bP[S_q = i] f_{Y_q \mid H, S_q}(\vy_q \mid \mH, i) \\
    &= c_0 \rho_i f_{Z_q}(\vy_q - \mH\vx_i),
\end{align*}
where $c_0$ is normalization constant independent of $i$. Thus, for a constant $c$ independent of $i$, 
\begin{align*}
    &f_{Z_q}(\vy_q - \mH\vx_i) \\
    &= c \exp\left(-\frac{1}{2}(\vy_q-\mH\vx_i)^T \bSigma_z^{-1}(\vy_q-\mH\vx_i)\right).
\end{align*}
Simplifying we get
\begin{align*}
    &(\vy_q-\mH\vx_i)^T \bSigma_z^{-1}(\vy_q-\mH\vx_i) \\
    &= \vy_q^T \bSigma_z^{-1} \vy_q + \vx_i^T \mH^T \bSigma_z^{-1} \mH \vx_i - 2 \vy^T_q \bSigma^{-1} \mH\vx_i.
\end{align*}
Further, denoting $\vh_I \triangleq \operatorname{Re}(\tilde{\vh}), \vh_Q \triangleq \operatorname{Im}(\tilde{\vh})$
\begin{align*}
    &\frac{1}{2} \mH^T \bSigma_z^{-1} \mH \\
    &=  \begin{bmatrix}
        \vh_I^T & \vh_Q^T \\
        -\vh_Q^T & \vh_I^T
    \end{bmatrix}
    \begin{bmatrix}
    \tilde{\bSigma}_z^{-1} & \0 \\
    \0& \tilde{\bSigma}_z^{-1}
\end{bmatrix}
 \begin{bmatrix}
        \vh_I & -\vh_Q \\
        \vh_Q & \vh_I
    \end{bmatrix} \\
    &=
    \begin{bmatrix}
        \vh_I^T & \vh_Q^T \\
        -\vh_Q^T & \vh_I^T
    \end{bmatrix}
    \begin{bmatrix}
        \tilde{\bSigma}_z^{-1}\vh_I & -\tilde{\bSigma}_z^{-1}\vh_Q \\
        \tilde{\bSigma}_z^{-1}\vh_Q & \tilde{\bSigma}_z^{-1}\vh_I
    \end{bmatrix}\\
    &=
    \begin{bmatrix}
       \vh_I^T \tilde{\bSigma}_z^{-1} \vh_I + \vh_Q^T \tilde{\bSigma}_z^{-1} \vh_Q & 0 \\
       0 & \vh_I^T \tilde{\bSigma}_z^{-1} \vh_I + \vh_Q^T \tilde{\bSigma}_z^{-1} \vh_Q
    \end{bmatrix} \\
    &\triangleq \gamma_{z, \vh} \mI_2,
\end{align*}
where $\gamma_{z, \vh} \triangleq \vh_I^T \tilde{\bSigma}_z^{-1} \vh_I + \vh_Q^T \tilde{\bSigma}_z^{-1} \vh_Q$ denotes the instantaneous signal to noise ratio (SNR). Therefore, using $\vx_i^T \vx_i = 1$ for all $i$, we have that 
\begin{align*}
    \frac{1}{2}\vx_i^T \mH^T \bSigma^{-1} \mH \vx_i = \gamma_{z,\vh},
\end{align*}
and so we obtain after the cancellation of constant terms,
\begin{align*}
    &\bP[S_q = i \mid H=\mH, Y_q=\vy_q] \\
    &= \frac{\rho_i \exp(\vy_q^T \bSigma_z^{-1}\mH \vx_i)}{\sum_{j\in[S]}\rho_j \exp(\vy_q^T \bSigma_z^{-1}\mH \vx_j)}.
\end{align*}

\emph{Proof of Lemma \ref{lem:convergence-SA}}. We can write the estimator as
\begin{align*}
    p^N_i(\mU_N; \mW) = \frac{ \left(\frac{N_i}{N} \right) \exp(\vy_q^T \mW \mH \vx_i) \left(\frac{1}{N_i} \sum_{n\in \cS_i}\exp(\vy_q^T \mW \vz_n) \right)}{\sum_{j\in[S]}  \left(\frac{N_j}{N} \right) \exp(\vy_q^T \mW \mH \vx_j) \left(\frac{1}{N_j} \sum_{m\in \cS_j}\exp(\vy_q^T \mW \vz_m) \right)},
\end{align*}
where $N_i = \abs{\cS_i(N)} \triangleq \{n\in[N] \mid s_n = i\} = \sum_{n\in[N]} \bI[s_n = i]$ is a binomial random variable with parameters $(N, \rho_i)$. Hence, by the strong law of large numbers, $N_i / N \to \rho_i$ for $i\in[S]$ a.s. On this almost sure set, $N_i \to \infty$, and so by another application of strong law of large numbers to the i.i.d. log-normal random variables with the finite (fourth) moment, almost surely, we have
\begin{align*}
    \frac{1}{\abs{\cS_i(N)}} \sum_{n\in \cS_i(N)}\exp(\vy_q^T \mW \vz_n)  \to  \bE_{\vz_0}[\exp(\vy_q^T \mW \vz_0)] > 0, \quad i\in[S].
\end{align*}
Thus, we have almost surely
\begin{align*}
     \lim_{N\to \infty} p^N_i(\mU_N; \mW) = \frac{\rho_i \exp(\vy_q^T \mW \mH \vx_i)}{\sum_{j\in[S]}\rho_j \exp(\vy_q^T \mW \mH \vx_j)} = p_i(\vy_q, \mH; \mW).
\end{align*}

\emph{Proof of Lemma \ref{lem:convexity-of-log-sum-exp}}. Directly computing the gradient of $\ell$, we get
\begin{align*}
    \frac{\partial \ell(\vu)}{\partial u_i} = \frac{\exp(u_i)}{\sum_{j\in [S]} \exp(u_j)}.
\end{align*}

Thus, we have for any $\vu^1, \vu^2$
\begin{align*}
    &\ell(\vu^2)-\ell(\vu^1) - \langle \nabla \ell (\vu^1), \vu^2-\vu^1 \rangle \\
    &= -\log \left( \frac{\sum_{i\in[S]} e^{u^1_i}}{\sum_{j\in[S]} e^{u^2_j}} \right) + \frac{\sum_{i\in[S]} (u^1_i-u^2_i) e^{u^1_i}}{\sum_{j\in[S]} e^{u^1_j}} \\
    &= \frac{\sum_{i\in[S]} e^{u^1_i} \log \left(\frac{e^{u^1_i}}{e^{u^2_i}}\right)}{\sum_{j\in[S]} e^{u^1_j}} -\log \left( \frac{\sum_{i\in[S]} e^{u^1_i}}{\sum_{j\in[S]} e^{u^2_j}} \right) \ge 0,
\end{align*}
due to the log-sum inequality, with equality if and only if $\vu^1 = \vu^2 + c \mathbf{1}$ for some $c \in \bR$. If $\vv_1, \vv_2\in V$, and $\lambda \in [0,1]$, we have
\begin{align*}
    &(\ell \circ \vu)(\lambda \vv_1 + (1-\lambda) \vv_2) = \ell(\lambda \vu(\vv_1) + (1-\lambda) \vu(\vv_2)) \\
    &\le \lambda (\ell \circ \vu)(\vv_1) + (1-\lambda) (\ell \circ \vu)(\vv_2),
\end{align*}
and hence the result. $\hfill \square$

\section{Single Layer Experiment}

\begin{figure}[!ht]
    \centering
    \includegraphics[width=0.5\linewidth]{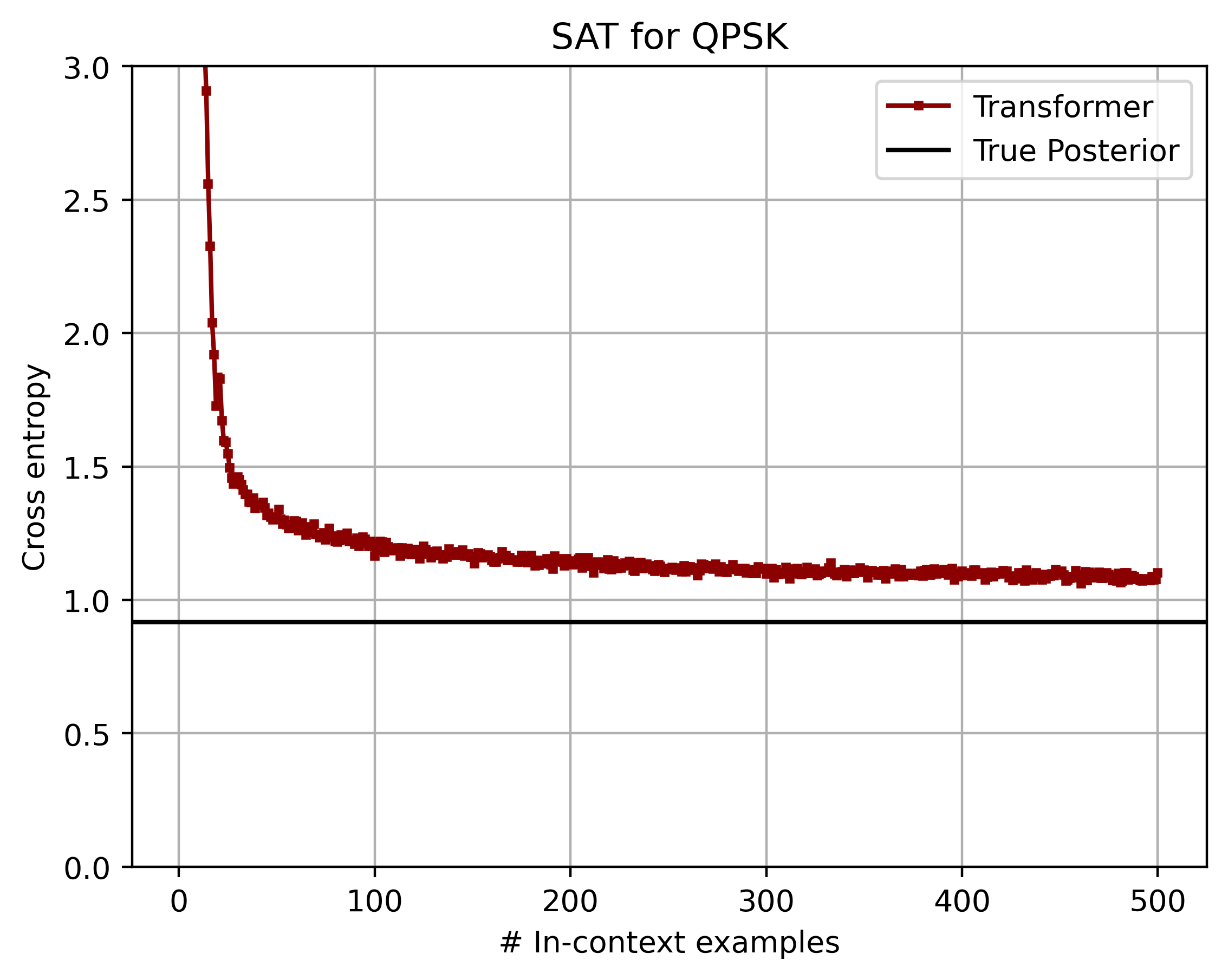}
    \caption{The performance of a single layer SAT approaches that of the true posterior as the number of in-context examples increases.}
    \label{fig:SAT-performance}
\end{figure}

In this section, we present a simple model trained and tested as the number of examples increases, in Figure \ref{fig:SAT-performance}. We train a single-layer softmax attention model as in the theoretical results with the trainable parameter $\mW$. We use the context length during training as $N=700$ and train for $1000$ epochs with a batch size of $128$. We use $d=4$ and the signal set of size $S = 4$ on the unit circle. We evaluate the posterior generated by the transformer using cross-entropy with the true symbol. We plot the cross-entropy of the true posterior as given by the Lemma \ref{lem:true-posterior} as the black horizontal line. As the number of examples during testing increases, the performance of the transformer approaches that of the true posterior. Note that the convergence is slow as it involves sample means of log-normal random variables that have heavy-tails.

\section{Additional experiments}

\label{appen:accuracy-plots}

% In addition to the accuracy plots in the main paper, we plot the cross-entropy of the posterior estimate by the transformer and other methods, in Figures \ref{fig:CE-TI} and \ref{fig:CE-TV}. The cross-entropy is a measure of the quality of the posterior estimate, the lesser the better. This is meaningful as opposed to the raw accuracy as the input to the decoder as a typical communication system are the posterior estimates. 

% \begin{figure}[!ht]
%     \centering
%     \includegraphics[width=\linewidth]{figures/CE_snr_neg5_QPSK_time_invariant.png}
%     \caption{Cross entropy for time-invariant process for QPSK signal set}
%     \label{fig:CE-TI}
% \end{figure}

% \begin{figure}
%     \centering
%     \includegraphics[width=\linewidth]{figures/CE-time-variant-SNR0-baselines-QPSK.png}
%     \caption{Cross entropy for time-varying process for QPSK signal set}
%     \label{fig:CE-TV}
% \end{figure}

\begin{figure}[!ht]
    \centering
    \includegraphics[width=\linewidth]{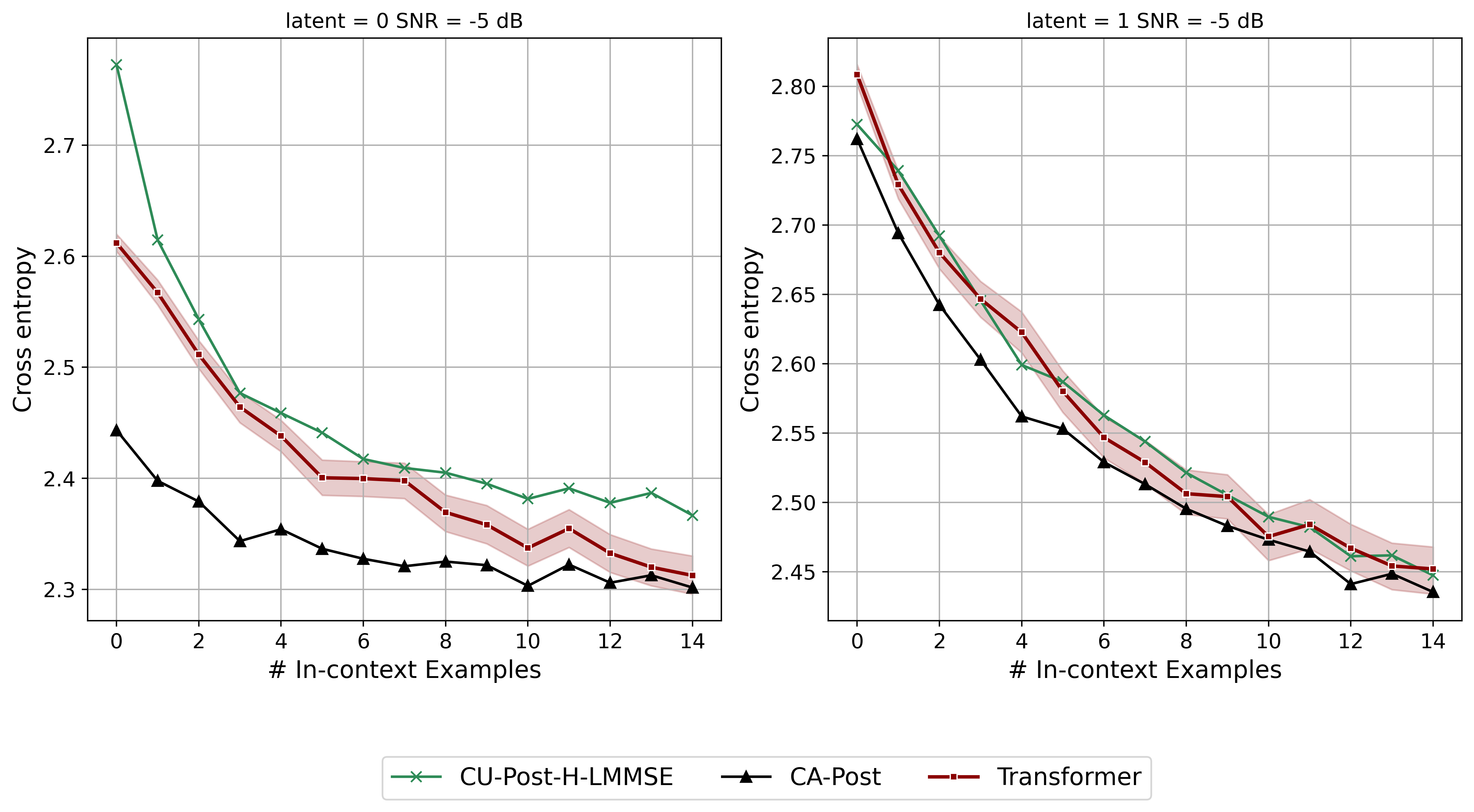}
    \caption{Cross entropy for time-invariant process for 16-QAM signal set}
    \label{fig:QAM-time-invariant}
\end{figure}

\begin{figure}[!ht]
    \centering
    \includegraphics[width=\linewidth]{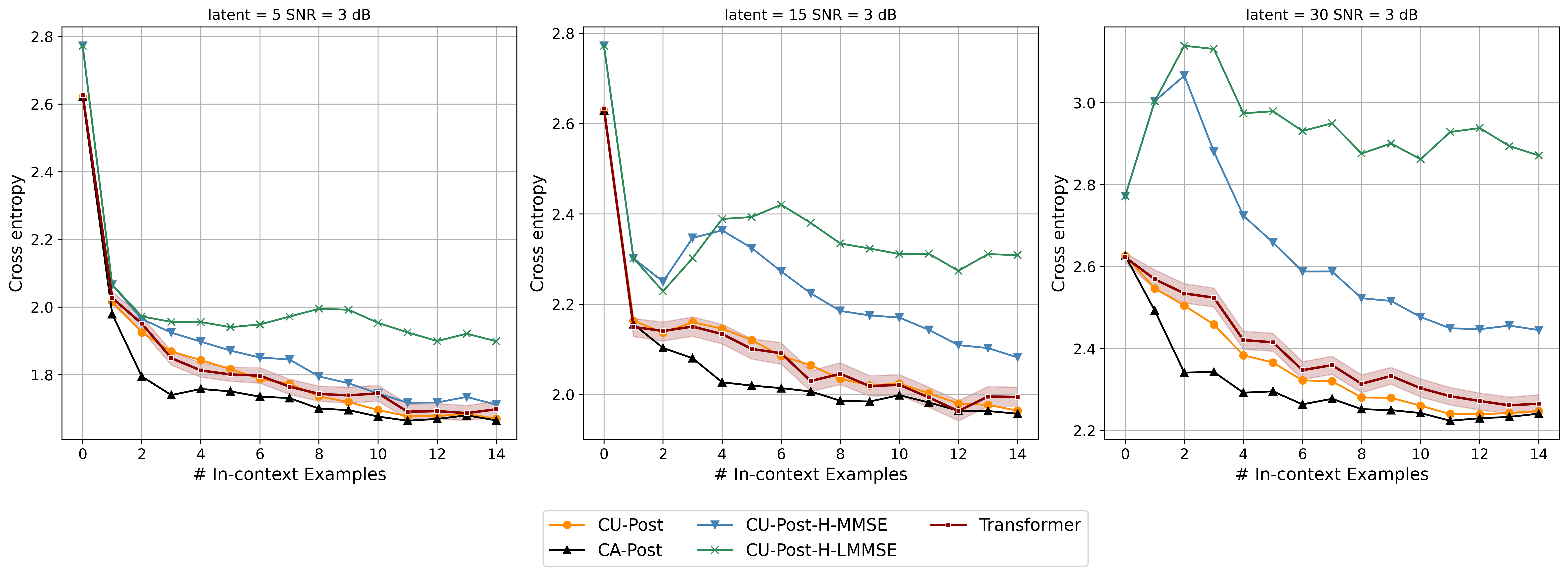}
    \caption{Cross entropy for time-varying process for 16-QAM signal set}
    \label{fig:QAM-time-variant}
\end{figure}

In this section, we provide the results with one more signal set, known as 16-QAM (Quadrature Amplitude Modulation). The points are evenly spaced in a grid of $4\times4$ and normalized to have average power $1$. Similar to the original setting the transformer performance is better than typical baselines and approaches to that of the true context-aware upper bound as shown in Figures \ref{fig:QAM-time-invariant} and  \ref{fig:QAM-time-variant}.

\begin{figure}[!ht]
    \centering
    \includegraphics[width=\linewidth]{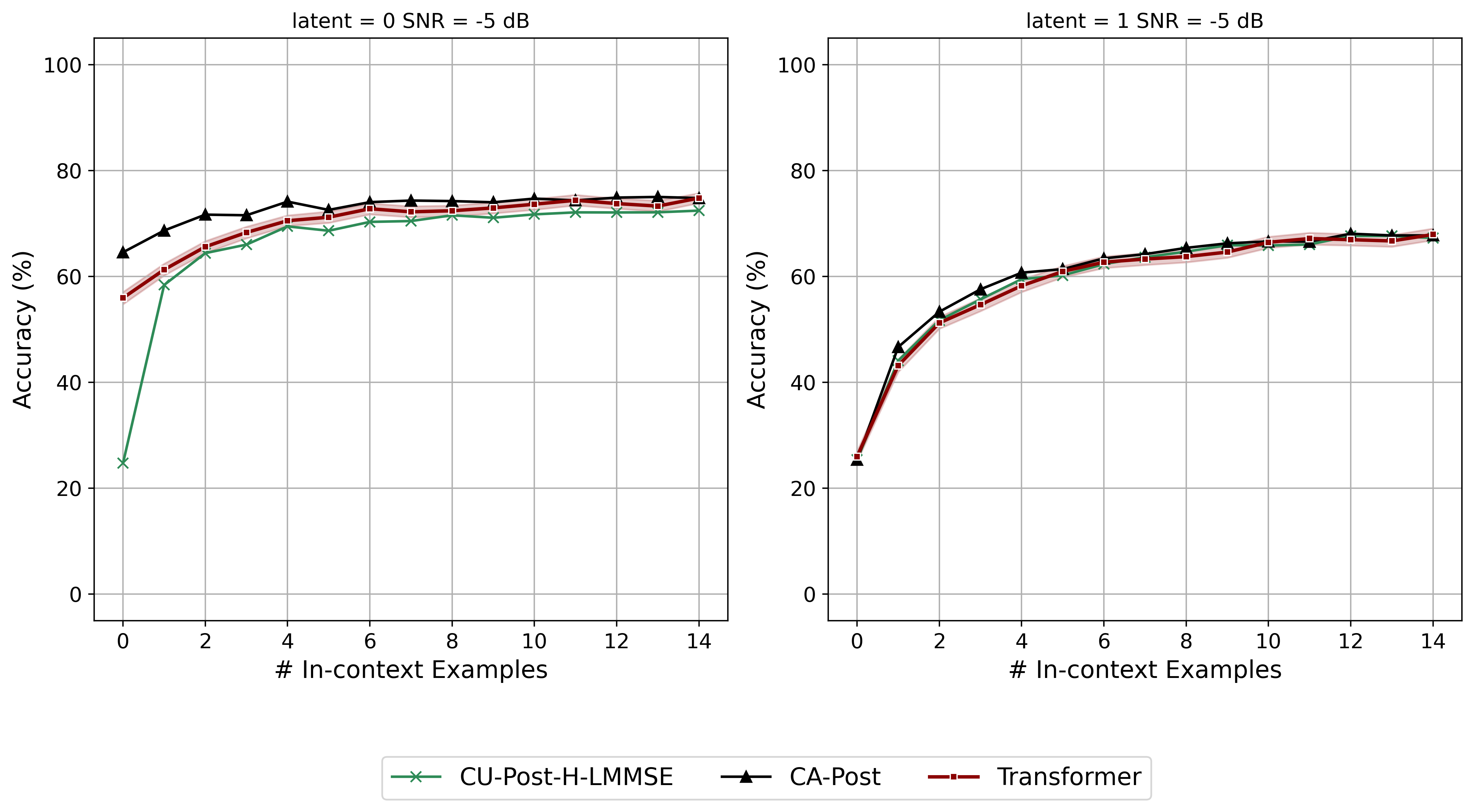}
    \caption{Accuracy plots for time invariant process}
    \label{fig:accuracy-time-invariant-snr-neg5-qpsk}
\end{figure}

\begin{figure}
    \centering
    \includegraphics[width=\linewidth]{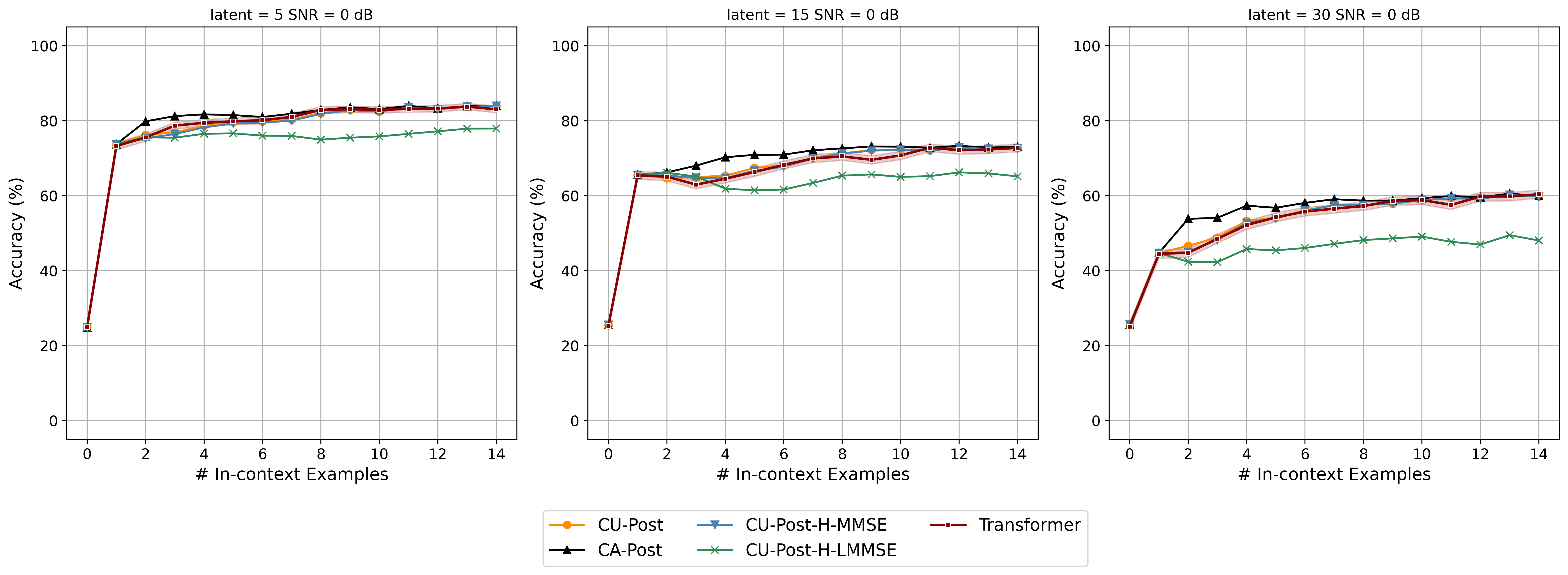}
    \caption{Accuracy plots for time variant process}
    \label{fig:accuracy-time-variant-snr-0-qpsk}
\end{figure}

Finally, we provide the plots for accuracy (win-rate) of predicting the current symbol from the estimated posterior distribution for scenario 1 (in Figure \ref{fig:accuracy-time-invariant-snr-neg5-qpsk}) and scenario 2 (in Figure \ref{fig:accuracy-time-variant-snr-0-qpsk}). Given an estimate for the distribution $\hat{p}_{S_q}$, the estimate of the symbol is given by $\hat{s}_q = \arg\max_{i\in[S]} \hat{p}_{S_q}(i)$ (this is the maximum aposteriori estimate). The accuracy is computed as ${\rm acc}(\%) \triangleq \frac{100}{N_{\rm stat}}\sum_{j\in [N_{\rm stat}]}\mathbf{1}_{\{\hat{s}_q^{(j)} = s_q^{(j)}\}}$, where $N_{\rm stat}$ is the number of statistical samples used in evaluating the performance of the estimator. In our experiments $N_{\rm stat}=10000$. These plots show that lesser cross-entropy translates to higher accuracy.

\section{Ablation}

\begin{figure}[!ht]
    \centering
    \includegraphics[width=\linewidth]{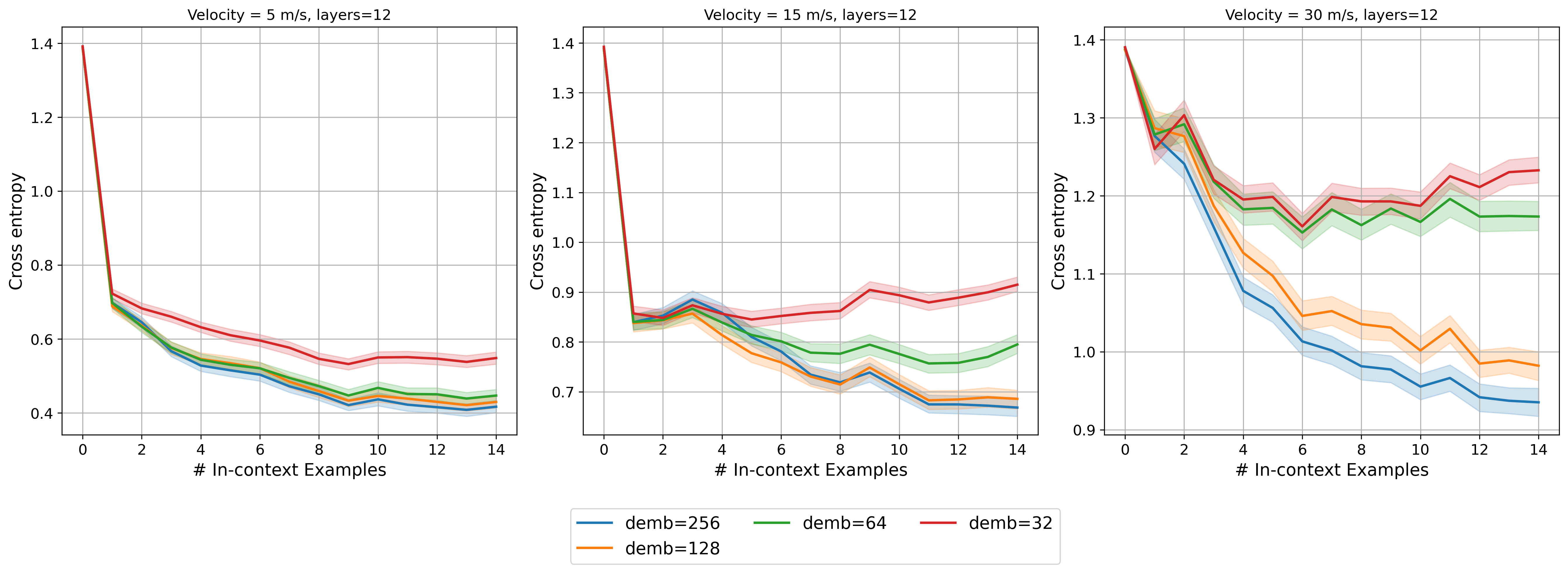}
    \caption{Effect of embedding dimension}
    \label{fig:ablation-demb}
\end{figure}

\begin{figure}
    \centering
    \includegraphics[width=\linewidth]{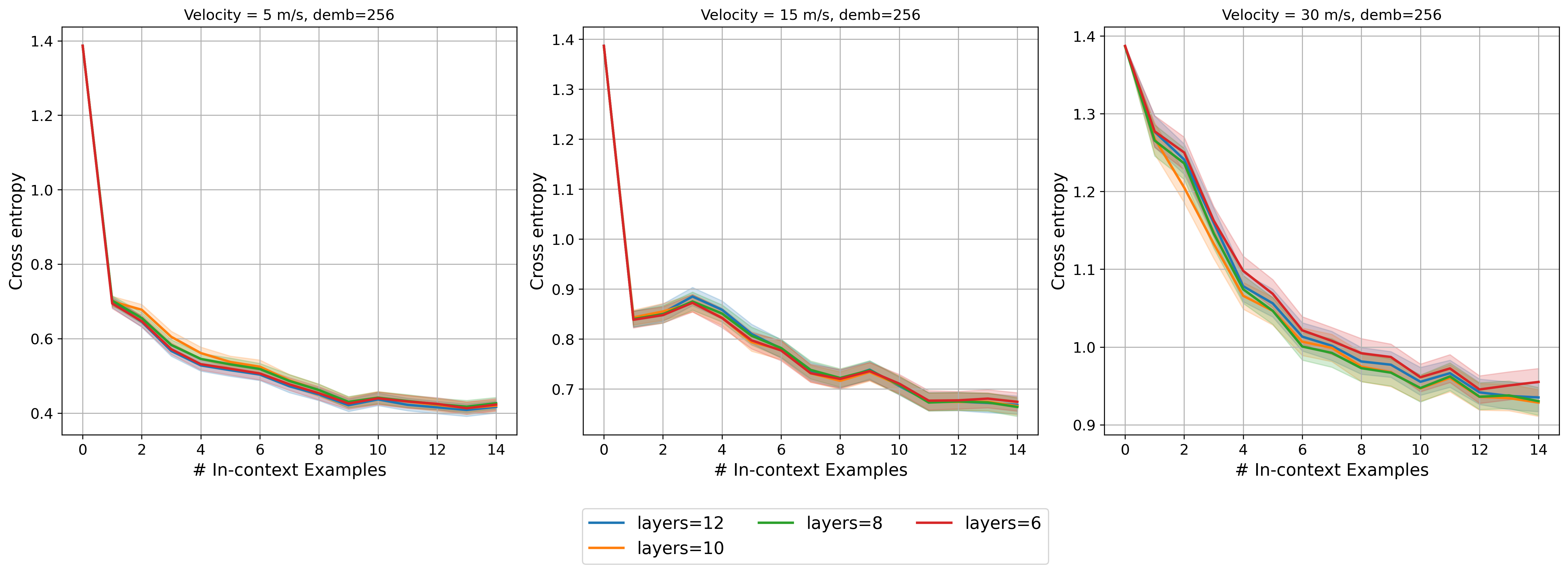}
    \caption{Effect of number of layers}
    \label{fig:ablation-nl}
\end{figure}

For some ablation results, see Figures \ref{fig:ablation-demb} and \ref{fig:ablation-nl}. The effect of embedding dimension is very significant to the performance of the transformer model. For the standard embedding size of $256$, the effect of layers down to $6$ is still not that significant.

\section{Derivation of Baselines}
\label{appen:baselines}

Consider the problem of estimating the symbol index $s_{k}$ from $\tilde{\vy}_0, \dots, \tilde{\vy}_{k-1}, \tilde{\vy}_{k}$ and $\tilde{x}_{s_0}, \dots, \tilde{x}_{s_{k-1}}$. 

For $\vx \in\bR^2$ corresponding to $\tilde{x}\in \bC$, define $\mM^d(\vx) \triangleq \begin{bmatrix}
    \operatorname{Re}(\tilde{x}) & -\operatorname{Im}(\tilde{x}) \\ \operatorname{Im}(\tilde{x}) & \operatorname{Re}(\tilde{x})
\end{bmatrix} \otimes \mI_d \in \R^{2d\times 2d}$, where $\otimes$ denotes the Kronecker product.

\subsection{Scenario 1}

The real equations are given by $\vy_n = \mM^d(\vx_{s_n}) \vh(\theta) + \vz_n$ for $n\in [k+1]$. Let $\vy_{\rm past} \triangleq [\vy_0^T, \dots, \vy_{k-1}^T]^T \in \bR^{2dk}$ denote the vector of past output symbols, and $\mX_{\rm past} \triangleq \begin{bmatrix} \mM^d(\vx_{s_0}) \\
\vdots \\
\mM^d(\vx_{s_{k-1}})
\end{bmatrix} \in \bR^{2dk \times 2d}$. Let $\vy_{\rm full} = [\vy_{\rm past}^T, \vy_{k+1}^T]^T\in \bR^{2d(k+1)}$ and $\mX_{\rm full}(i) \triangleq \begin{bmatrix}
    \mX_{\rm past} \\
    \mM^d(\vx_i)
\end{bmatrix} \in \bR^{2d(k+1) \times 2d}$ for each $i \in [S]$, $\vh(\theta) = [\vh_I^T,~ \vh_Q^T]^T \in \bR^{2d}$. Thus, we have the problem of estimating $s_{k}$ using $\vy_{\rm full} = \mX_{\rm full}(s_k) \vh(\theta) + \vz_{\rm full}$, where $\vz_{\rm full} \in \bR^{2d(k+1)}$ is defined similar to $\vy_{\rm full}$. We compute the posterior as
\begin{align*}
    &\bP[s_{k+1} = i \mid Y_{\rm full} = \vy_{\rm full}, X_{\rm past} = \mX_{\rm past}] \\
    &\propto f_{Y_{\rm full}, X_{\rm past}, S_{k+1}}(\vy_{\rm full}, \mX_{\rm past}, i) \\
    &= \rho_i f_{X_{\rm past}}(\mX_{\rm past}) f_{Y_{\rm full} \mid X_{\rm past}, S_{k+1}}(\vy_{\rm full} \mid \mX_{\rm past}, i) \\
    &\propto \rho_i f_{Y_{\rm full} \mid X_{\rm full}}(\vy_{\rm full} \mid \mX_{\rm full}(i)).
\end{align*}
Now, we compute the overall likelihood density as
\begin{align*}
      &f_{Y_{\rm full} \mid X_{\rm full}}(\vy_{\rm full} \mid \mX_{\rm full}(i)) \\
      &= \int_{\theta} f_{Y_{\rm full},  \Theta \mid X_{\rm full}}(\vy_{\rm full}, \theta \mid \mX_{\rm full}(i)) d\theta \\
      &= \int_{\theta} f_\Theta(\theta) f_{Y_{\rm full} \mid X_{\rm full}, \Theta}(\vy_{\rm full} \mid \mX_{\rm full}(i), \theta) d\theta,
\end{align*}
where the second equality follows from $\theta$ being independent of $\mX_{\rm full}(i)$.

For scenario 1, the latent space $\{0,1\}$ is finite. Thus, we have
\begin{align*}
    &\ell_0(i) \triangleq f_{Y_{\rm full} \mid X_{\rm full}, \Theta}(\vy_{\rm full} \mid \mX_{\rm full}(i), 0) \\
    &= \int_{\alpha \in (0, \pi]} f_{A, Y_{\rm full} \mid X_{\rm full}, \Theta}(\alpha, \vy_{\rm full} \mid \mX_{\rm full}(i), 0) d\alpha \\
    &= \int_{\alpha \in (0, \pi]} f_A(\alpha) f_{Y_{\rm full} \mid X_{\rm full}, \Theta, A}(\vy_{\rm full} \mid \mX_{\rm full}(i), 0, \alpha) d\alpha \\
    &= \int_{\alpha \in (0, \pi]} f_A(\alpha) f_{Y_{\rm full} \mid X_{\rm full}, \tilde{H}}(\vy_{\rm full} \mid \mX_{\rm full}(i), \tilde{\vh}_{\alpha}(0)) d\alpha \\
    &= \int_{\alpha \in (0, \pi]} f_A(\alpha) f_{\tilde{Y}_k \mid \tilde{H}, S_{k}}(\tilde{\vy}_k \mid \tilde{\vh}_{\alpha}(0), i) \prod_{n=0}^{k-1} f_{\tilde{Y}_n \mid \tilde{H}, S_{n}}(\tilde{\vy}_n \mid \tilde{\vh}_{\alpha}(0), s_{n}) d\alpha \\
    % &= \int_{\alpha \in (0, \pi]} p(\alpha) p_{\tilde{\vz}}(\tilde{\vy}_{k+1}-\tilde{x}\tilde{\vh}_{\alpha}(0))  \prod_{i=1}^{k} p_{\tilde{\vz}}(\tilde{\vy}_{i}-\tilde{x}_i\tilde{\vh}_{\alpha}(0)) d\alpha \\
    &= \frac{1}{(2\pi)^{d(k+1)} (\sigma^{2d(k+1)})} \int_{\alpha \in (0, \pi]} f_A(\alpha) \exp\left(-\norm{\tilde{\vy}_{k}-\tilde{x}_i\tilde{\vh}_{\alpha}(0)}_2^2/(2\sigma^2)\right)  \prod_{n=0}^{k-1} \exp\left(-\norm{\tilde{\vy}_{n}-\tilde{x}_{s_n}\tilde{\vh}_{\alpha}(0)}_2^2/(2\sigma^2)\right) d\alpha,
\end{align*}
where $f_A(\alpha) = \frac{1}{\pi}$ and $\tilde{\vh}_\alpha(0)$ denotes the dependence of the hidden parameter on $\alpha$. The above one-dimensional integral can be computed numerically.

Conditioned on $\mX_{\rm full}(i), \theta=1$, since $\vy_{\rm full} = \mX_{\rm full}(i) \vh(1) + \vz_{\rm full}$ is Gaussian random variable with mean $\0$ and covariance $\mC(i) \triangleq 1/2 \cdot \mX_{\rm full}(i) \mX_{\rm full}^T(i) + \sigma^2 \mI_{2d(k+1)}$, we have 
\begin{align*}
    \ell_1(i)\triangleq f_{Y_{\rm full}\mid X_{\rm full}, \Theta}(\vy_{\rm full} \mid  & \mX_{\rm full}(i), 1) = \frac{1}{(2\pi)^{d(k+1)}\sqrt{\abs{\det(\mC(i))}}} \exp(-1/2 \cdot \vy_{\rm full}^T \mC^{-1}(i) \vy_{\rm full}).
\end{align*}

Thus $\bP[s_k = i \mid Y_{\rm full} = \vy_{\rm full}, X_{\rm past} = \mX_{\rm past}] \propto \rho_i (f_\Theta(0) \ell_0(i) + f_\Theta(1) \ell_1(i))$. If the context is known, then the context-aware estimator is given by
\begin{align}
\label{eq:scenario-1-aware}
    \bP[s_k = i \mid Y_{\rm full} = \vy_{\rm full}, X_{\rm past} = \mX_{\rm past}, \Theta = \theta] \propto \rho_i \ell_\theta(i).
\end{align}

\subsection{Scenario 2}

The real equations are given by $\vy_n= \mM^d(\vx_{s_n}) \vh_n(\theta) + \vz_n$ for $n \in [k+1]$. Let $\vy_{\rm past} \triangleq [\vy_1^T, \dots, \vy_k^T]^T \in \bR^{2dk}$ to be the vector of past output symbols, and $\mX_{\rm past} \triangleq {\rm diag}(\mM^d(\vx_{s_0}), \dots, \mM^d(\vx_{s_{k-1}}))\in \bR^{2dk \times 2dk}$ to be the block diagonal matrix from past examples. Let $\vy_{\rm full} \triangleq [\vy_{\rm past}^T, \vy_{k+1}^T]^T\in \bR^{2d(k+1)}$ and $\mX_{\rm full}(s_{k}) \triangleq {\rm diag}(\mX_{\rm past}, \mM^d(\vx_{s_{k}})) \in \bR^{2d(k+1) \times 2d(k+1)}$. Then, the equations are given by $\vy_{\rm full} = \mX_{\rm full}(\vx_{s_k}) \vh_{\rm full}(\theta) + \vz_{\rm full}$, where $\vh_{\rm full}(\theta), \vz_{\rm full} \in \bR^{2d(k+1)}$ are defined similar to $\vy_{\rm full}$. We can write the context-unaware posterior estimate as
\begin{align*}
    \bP[s_{k} = i \mid Y_{\rm full} = \vy_{\rm full}, X_{\rm past} = \mX_{\rm past}] &\propto \rho_i f_{Y_{\rm full} \mid X_{\rm full}} (\vy_{\rm full} \mid \mX_{\rm full}(i))  \\
    &= \rho_i \left(\int_{\theta} f_\Theta(\theta) f_{Y_{\rm full} \mid X_{\rm full}, \Theta} (\vy_{\rm full} \mid \mX_{\rm full}(i), \theta) d\theta \right),
\end{align*}
similar to that in the previous section. Hence, it suffices to compute $f_{Y_{\rm full} \mid X_{\rm full}, \Theta} (\vy_{\rm full} \mid \mX_{\rm full}(i), \theta)$ for each $\theta$. In scenario 2, note that $\vh_{\rm full}(\theta)$ given $\theta$ is a Gaussian random vector with correlation $\mR_{\rm full}(\theta) \triangleq \bE[\vh_{\rm full}(\theta)\vh_{\rm full}^T(\theta)] = \mR_\theta \otimes \mI_{2d}$, where $(\mR_\theta)_{i,j} = R_\theta(\abs{i-j})$ for $i,j\in[k+1]$, where $R_\theta(\cdot)$ is defined previously. Given $\mX_{\rm full}(i), \theta$, the vector $\vy_{\rm full}$ can be seen to be a Gaussian random vector with zero mean and covariance $\mC_\theta(i) \triangleq  \mX_{\rm full}(i) \mR_{\rm full}(\theta) \mX_{\rm full}^T(i) + \sigma^2 \mI_{2d(k+1)}$. Therefore, we get that
\begin{align*}
    \ell_\theta(i) \triangleq f_{Y_{\rm full} \mid X_{\rm full}, \Theta} (\vy_{\rm full} \mid \mX_{\rm full}(i), \theta) \propto \frac{1}{\sqrt{\abs{\det(\mC_\theta(i))}}} \exp(-1/2 \cdot \vy_{\rm full}^T \mC_\theta^{-1}(i) \vy_{\rm full}).
\end{align*}
If the latent space of $\theta$ is finite, we can compute the posterior distribution as
\begin{align*}
    \bP[s_{k} = i \mid Y_{\rm full} = \vy_{\rm full}, X_{\rm past} = \mX_{\rm past}] \propto \rho_i \sum_{\theta} f_\Theta(\theta) \ell_\theta(i).
\end{align*}
Similar to previous section, the context-aware posterior estimate can be computed as
\begin{align*}
    \bP[s_{k} = i \mid Y_{\rm full} = \vy_{\rm full}, X_{\rm past} = \mX_{\rm past}, \Theta=\theta] \propto \rho_i f_\Theta(\theta) \ell_\theta(i).
\end{align*}

\subsection{Typical baselines}

\label{appen:typical-baselines}

We describe CU-Post-H-EST for ${\rm EST} = {\rm MMSE}, {\rm LMMSE}$. Typically one computes the estimate $\hat{\vh}^{\rm EST}$  of $\vh_{k}$ from $\vy_{\rm past}$ and then compute the posterior as 

\begin{align*}
    \bP[s_k = i \mid Y_k = \vy_k, H_k = \vh_{k}^{\rm EST}] \propto \rho_i f_Z(\vy_{k}-\mH_k^{\rm EST} \vx_i) \propto \exp(-\norm{\vy_k-\mH_k^{\rm EST}\vx_i}_2^2/(2\sigma^2)).
\end{align*}

% where $\bP_x(\vx_{k+1} = \vx \mid \vy_{k+1}, \vh_{k+1} = \vh_{k+1}^{\rm EST}) \propto \bP_x(\vx) \bP(\vy_{k+1} \mid \vx_{k+1} = \vx, \vh_{k+1} = \vh_{k+1}^{\rm EST})$ and we have
% \begin{align*}
%     \bP(\vy_{k+1} \mid \vx_{k+1} = \vx, \vh_{k+1} = \vh_{k+1}^{\rm EST}) = p_\vz(\vy_{k+1}-\mH_{k+1}^{\rm EST} \vx) \propto \exp(-\norm{\vy_{k+1}-\mH_{k+1}^{\rm EST} \vx}_2^2/(2\sigma^2)).
% \end{align*}

Now we show how to compute $\vh_{k+1}^{\rm EST}$. The MMSE estimate can be computed as

\begin{align*}
    \vh_{k}^{\rm MMSE} &= \bE[\vh_{k} \mid Y_{\rm past} = \vy_{\rm past}, X_{\rm past} = \mX_{\rm past}] \\
    &= \bE[\bE[\vh_{k} \mid Y_{\rm past} = \vy_{\rm past}, X_{\rm past} = \mX_{\rm past}, \Theta=\theta] \mid Y_{\rm past} = \vy_{\rm past}, X_{\rm past} = \mX_{\rm past}] \\
    &= \bE[ \vh_{k}^{\rm MMSE, \theta} \mid  Y_{\rm past} = \vy_{\rm past}, X_{\rm past} = \mX_{\rm past}] \\
    &= \frac{\int_\theta f_\Theta(\theta) f_{Y_{\rm past} \mid X_{\rm past}, \Theta}(\vy_{\rm past} \mid \mX_{\rm past}, \theta) \vh_{k}^{\rm MMSE, \theta} d\theta}{\int_\theta f_\Theta(\theta) f_{Y_{\rm past} \mid X_{\rm past}, \Theta}(\vy_{\rm past} \mid \mX_{\rm past}, \theta) d\theta}.
\end{align*}

% \begin{align*}
%     &\vh_{k+1}^{\rm MMSE} = \bE[\vh_{k+1} \mid \vy_{\rm past}, \mX_{\rm past}] \\
%     &= \int_{\vh} \vh p(\vh_{k+1} = \vh \mid \vy_{\rm past}, \mX_{\rm past}) d\vh = \int_{\theta \in \Theta} \int_{\vh} \vh p(\vh_{k+1} =\vh, \theta \mid \vy_{\rm past}, \mX_{\rm past}) d\vh d\theta \\
%     &= \int_{\theta \in \Theta} p(\theta \mid \vy_{\rm past}, \mX_{\rm past})  \int_{\vh} \vh p(\vh_{k+1} =\vh \mid \vy_{\rm past}, \mX_{\rm past}, \theta) d\vh d\theta \\
%     &= \int_{\theta \in \Theta} p(\theta \mid \vy_{\rm past}, \mX_{\rm past})  \bE[\vh_{k+1} \mid \vy_{\rm past}, \mX_{\rm past}, \theta] d\theta \\
%     &= \int_{\theta \in \Theta} p(\theta \mid \vy_{\rm past}, \mX_{\rm past}) \vh_{k+1}^{\rm MMSE, \theta} d\theta = \frac{\int_{\theta \in \Theta} p(\theta) p(\vy_{\rm past} \mid \mX_{\rm past}, \theta)  \vh_{k+1}^{\rm MMSE, \theta} d\theta}{\int_{\theta \in \Theta} p(\theta) p(\vy_{\rm past} \mid \mX_{\rm past}, \theta) d\theta}.
% \end{align*}

Since given $\theta$, $\vy_{\rm past}$ and $\vh_{k}$ are jointly Gaussian MMSE of $\vh_{k}$ can be computed using LMMSE of $\vh_{k}$ from $\vy_{\rm past}$. First, we have $\mR_{h,y}(\theta) \triangleq \bE[\vh_{k}(\theta) \vy_{\rm past}^T] \in \bR^{2d \times 2dk}$, where $\mR_{h,y}(\theta) = (\mR_\theta[k, 0:k-1] \otimes \mI_{2d}) \mX_{\rm past}^T$, where $[\mR_\theta]_{i,j} = R_\theta(\abs{i-j})$; we used the Python notation. Also, $\mR_{y,y}(\theta) \triangleq \mX_{\rm past} (\mR_\theta[0:k-1, 0:k-1] \otimes \mI_{2d}) \mX_{\rm past}^T + \sigma^2 \mI_{2dk}$. Thus, can compute $\vh_{k}^{\rm MMSE, \theta} = \mR_{h,y}(\theta) \mR_{y,y}^{-1}(\theta) \vy_{\rm past}$. When latent space of $\theta$ is finite, overall $\vh_{k+1}^{\rm MMSE}$ can be computed from the last equation. Finally, $f_{Y_{\rm past} \mid X_{\rm past}, \Theta}(\vy_{\rm past} \mid \mX_{\rm past}, \theta)$ is the Gaussian density with mean $\0$ and covariance $\mR_{y,y}(\theta)$.

Lastly, one can compute the LMMSE estimate of the channel from $\vy_{\rm past}$ which is a very common practical algorithm. Let $\mR = \bE_{\theta\sim \Theta}[\mR_\theta]$ denote the average correlation of the hidden process  $\{\vh_n\}_{n\ge 0}$, which can be used in the previous computations to arrive at $\hat{\vh}_{k}^{\rm LMMSE} = \mR_{h,y} \mR_{y,y}^{-1} \vy_{\rm past}$ where the matrices $\mR_{h,y}, \mR_{y,y}$ are defined from the matrix $\mR$.

One can define the context-aware posterior estimate as done in the previous sections as
\begin{align*}
    \bP[s_k = i \mid Y_k = \vy_k, H_k = \vh_k^{\rm MMSE, \theta}].
\end{align*}

\subsection{Model Parameters}
\label{parameters}
We extend the codebase from~\cite{panwar2024incontext} for our models and experiments. We use the same GPT-2 model size as in this paper, as well as the same optimizer (Adam, \cite{Kingma2014AdamAM}), the same learning rate (0.0001), 
and the same training length (500,000 steps).

% \begin{table}[ht]
%     \centering
%     \begin{tabular}{|c|c|c|}
%          \hline
%          \textbf{Embedding Size} & \textbf{\# Layers} & \textbf{\# Attention Heads} \\
%          \hline
%          256 & 12 & 8\\
%          \hline
%     \end{tabular}
%     \vspace{2pt}
%     \caption{Model Parameters}
%     \label{tab:modelparameters}
% \end{table}

% \begin{table}[ht]
%     \centering
%     \begin{tabular}{|c|c|c|c|c|}
%          \hline
%          \textbf{Scenario}& \textbf{Curriculum Start Length} & \textbf{Curr. End Length} & \textbf{Batch Size} & \textbf{Learning Rate} \\
%          \hline
%          \textbf{C1} & 5 & 11 & 64 & 0.0001 \\
%          \hline
%          \textbf{C2} & 5 & 15 & 64 & 0.0001 \\
%          \hline
%     \end{tabular}
%     \vspace{2pt}
%     \caption{Training Parameters}
%     \label{tab:trainingparameters}
% \end{table}

\textbf{Curriculum:} \cite{garg2023transformers} observed that curriculum learning is useful in training models to perform in-context learning consistently and effectively.
In \cite{garg2023transformers} and \cite{panwar2024incontext}, the curriculum is to initiate training with small sequence lengths and vector dimensions and to increase these gradually over training until they reach the desired sizes. In our tasks, the vector dimension is fixed to twice the number of antennae,
and we modify only the sequence lengths in our curriculum. For scenario C1 and C2, we vary the sequence length from 5 to a maximum of 15. Batch size is chosen to be 64.

The training has been done using a single NVIDIA A100 GPU on an internal cluster. Training of a single transformer takes about 8 hours for standard model architecture. For ablation studies the embedding dimension is chosen to be 256, 128, 62, 32 with fixed 12 number of layers. On the other hand, for the embedding dimension of 256, number of layers was changed and trained for various models to be 10, 8, 6.

\section{Experiments with other neural sequence models}

\begin{figure}[!ht]
    \centering
    \includegraphics[width=\linewidth]{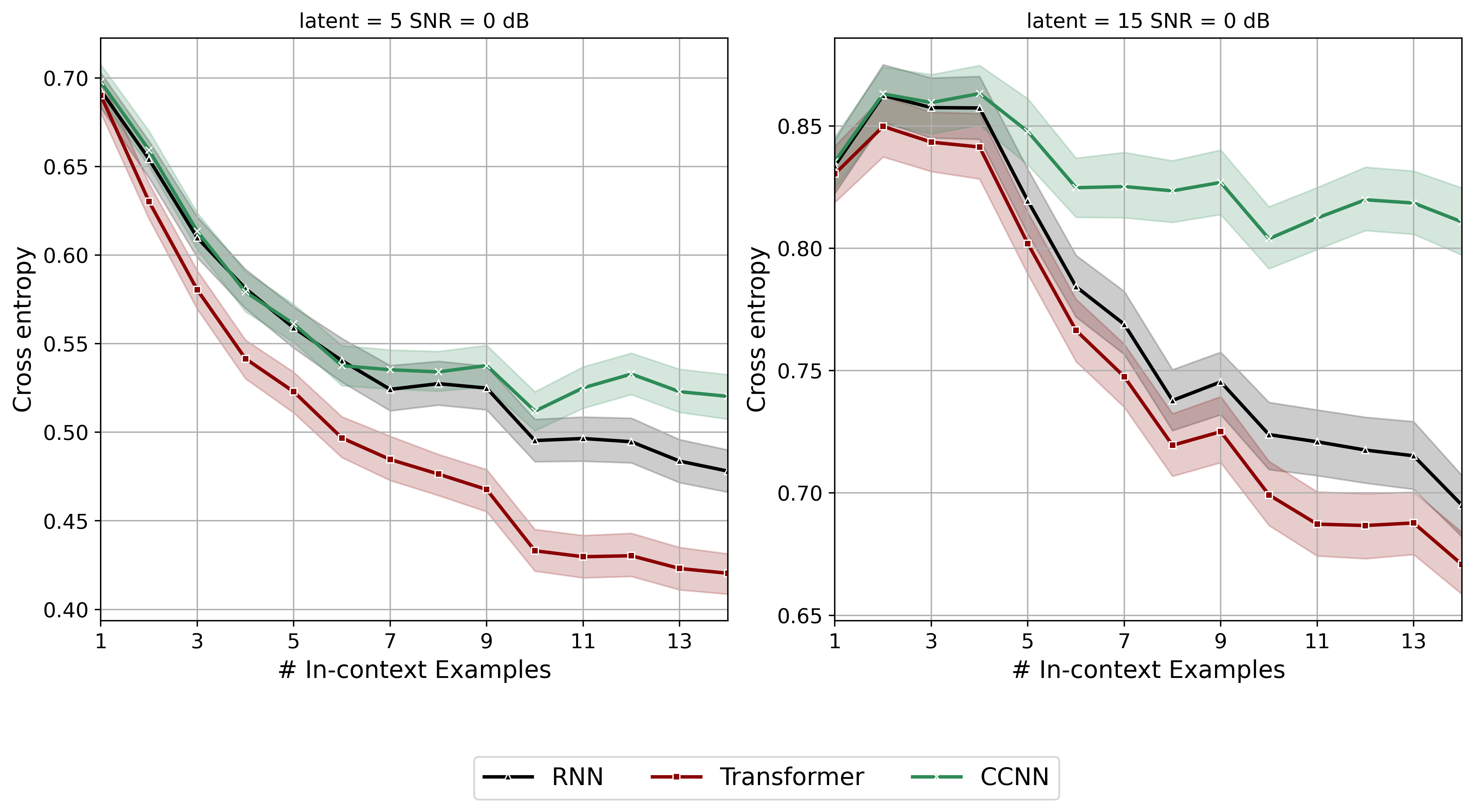}
    \caption{Performance of sequence models on the task of In-context Estimation in Scenario-2. The transformer performs uniformly better than the other models.}
    \label{fig:seq-models}
\end{figure}

We conduct the experiments with other sequence models like recurrent neural network (RNN) and causal convolutional neural network (CCNN). We tried various values for the number of layers and embedding dimension and used the best. For RNN, we used $4$ layers with embedding dimension of $512$. For CCNN, we used $6$ layers with embedding dimension of $256$, similar to the transformer. The RNN is implemented using the standard PyTorch module. CCNN is effectively 1D convolutions are applied to the zero-padded sequence with zeros only on one side. This preserves the causal structure of the problem. The performance comparison is reported in Figure \ref{fig:seq-models}. Clearly, transformers outperform both the models. In particular, CCNN has a plateaued behavior, 
 (possibly) indicating its inability to use the information effectively from the entire sequence in estimating the posterior distribution. RNN has the issue of generalizing for multiple contexts simultaneously, i.e., it performs well when latent is $15$ and fails when latent is $5$. Transformers, on the other hand, perform optimally on both the contexts, indicating better in-context estimation capabilities.

%% file: AISTATS-2025/contents-AISTATS/additional-related-work.tex
\section{Additional Related Work} 

\subsection{In context learning}

The in-context learning abilities of the transformer were first showcased by \cite{brown2020langmodelsfewshot} where GPT-3, a transformer-based large language model, was shown to perform well on novel, unseen tasks in a few-shot or zero-shot manner. The authors of \cite{xie2022explanation} used synthetic data to show that in-context learning can be interpreted as Bayesian inference, wherein a model first attempts to infer the task and then uses this prediction to carry out the task.

In \cite{garg2023transformers}, authors look at in-context learning as a problem of learning to perform linear regression by being trained on sequence data, rather than learning in the natural language setting. The authors of \cite{panwar2024incontext} build on this work and provide mathematical theory to describe how this can be interpreted as the model performing Bayesian inference by estimating a posterior distribution. These two works are closely related to our work.

\subsection{Theory of Attention Mechanisms}

Other work has gone further in explaining the mechanisms by which transformers are able to perform in-context learning. Some authors have explored the behavior of simplified transformer models on synthetic linear regression problems, compared to the globally optimal solutions for these problems\cite{mahankali2024onestep,zhang2024trained}.
In \cite{akyurek2023whatlearningalgorithm}, it is shown by construction that transformers can learn to fit a linear model at inference time given a context. This tells us that, at least in theory, transformers can learn to perform regression and related computations given some data in context. A related line of work explores how transformer models carry out these optimization procedures and shows that transformer models can perform nontrivial gradient-based optimization procedures at inference time\cite{von2023transformers,ahn2023transformers}. 
In \cite{chan2022datadist}, the authors test how properties of training data impact in-context learning behaviors. 
The authors of \cite{min2022rethinking} similarly study how the properties of the context itself impact in-context learning performance. 

In summary, many different research directions have provided ways to interpret and understand in-context learning and the mechanisms by which transformer models do it. Our work primarily uses the Bayesian perspective to view this ability and understand how we can apply it to the wireless communication domain.

% \subsection{Machine Learning for Wireless Communication}

\subsection{In context learning for wireless}

% The current literature is filled with machine-learning-based approaches to estimation in wireless communications, these mainly involve either symbol estimation or channel estimation. There are two main approaches to symbol estimation , one where first the channel is estimated and the symbol estimate is made conditioned on the channel estimate, the other approach is to make a direct symbol estimate. Our work takes the latter approach , while a significant portion of the literature is available on just channel estimation.  

The problem of symbol estimation in the presence of an unknown channel is a canonical problem in wireless communication and there is a large body of literature that considers model-driven traditional signal processing approaches. 
When the prior distribution for the channel follows a hierarchical model with latent contexts such as what we study in this paper, optimal model-based estimators become computationally infeasible. 
While approximations to the optimal estimators have been studied with restricted priors, the structure and performance of such estimators are highly dependent on the priors.

Recently, there has been significant interest in using machine learning methods such as variational inference \cite{baur2022variational}, \cite{caciularu2020channeleqvae} and deep neural network models such as fully connected neural networks \cite{DeepLearningAidedChannelEstimation}, convolutional neural networks \cite{LearningMMSEChannelEstimator}, and recurrent neural networks \cite{RNNforChannelEstimation} 
for channel estimation and symbol estimation \cite{aoudia2021end}. The authors of \cite{aoudia2021end} design an end-to-end wireless communication system design using a learning-based approach.
The work of \cite{LearningMMSEChannelEstimator} provides a principled way of designing channel estimators in the setting of wireless communication, by training a CNN-based neural network to efficiently estimate the channel parameters when it follows a hierarchical prior. The authors of \cite{caciularu2020channeleqvae} used variational autoencoders (VAEs) and unsupervised learning to implicitly learn and decode symbols in a wireless channel. In \cite{burshtein2023semi}, the authors use VAEs in a semi-supervised setting to operate over non-linear channels. 

% The approach in \cite{deepLearningForTimeVaryingMIMOChannels} uses a long short-term memory (LSTM) based channel parameter estimator for a time-varying channel. They are interested in learning the spatial correlation structure of the channel coefficients across the receiver antennae.
 
 The main distinction of our work from the previous works is that we make the connection between the in-context learning capabilities of the transformer and the symbol estimation problem in wireless communication. We show how the problem fits naturally in the setting, and provide theoretical and empirical evidence that transformers achieve near-optimal performance without needing additional input of channel realizations or estimates during training.